\documentclass[traditabstract]{aa} 

\usepackage{graphicx}
\usepackage{txfonts}
\usepackage{epstopdf}
\usepackage{subcaption}
\usepackage{color, colortbl}
\usepackage{bm}
\usepackage{array}
\usepackage{hyperref}
\usepackage{siunitx}
\usepackage{xcolor}

\newcommand{\NII}{[N{\scriptsize\,II}]}
\newcommand{\SII}{[S{\scriptsize\,II}]}
\newcommand{\CII}{[C{\scriptsize\,II}]}

\newcommand{\OI}{[O{\scriptsize\,I}]}
\newcommand{\HII}{H{\scriptsize\,II}}

\definecolor{Magenta}{rgb}{1,0,1}
\definecolor{LightGray}{gray}{0.9}

\begin{document} 

\title{Globules and pillars in Cygnus X}
\subtitle{IV. Velocity-resolved [OI] 63 $\mu$m map of a peculiar proplyd-like object}

\author{Nicola Schneider \inst{1} 
\and Simon Dannhauer \inst{1,2} 
\and Eduard Keilmann \inst{1} 
\and Slawa Kabanovic \inst{1} 
\and Theodoros Topkaras \inst{1} 
\and Volker Ossenkopf-Okada\inst{1} 
\and Ronan Higgins \inst{1}  
\and Andreas Brunthaler \inst{2} 
\and Won-Ju Kim \inst{1,2} 
\and Fernando Comer\'on \inst{3} 
\and Markus R\"ollig \inst{4,1} 
\and Timea Csengeri \inst{5} 
\and Robert Simon \inst{1} 
\and Yoko Okada \inst{1}  
\and Matthias Justen \inst{1} 
\and Sergio A. Dzib \inst{2}  
\and Gisela N. Ortiz-León \inst{6} 
}
 
\institute{I. Physikalisches Institut, Universität zu Köln, Z\"ulpicher Str. 77, 50937 K\"oln, Germany\\
\email{nschneid@ph1.uni-koeln.de}
\and Max-Planck Institut f\"ur Radioastronomie, Auf dem H\"ugel 69, 53121 Bonn, Germany 
\and European Southern Observatory, Karl-Schwarzschild-Straße 2, D-85748 Garching, Germany 
\and Physikalischer Verein, Gesellschaft für Bildung und Wissenschaft, Robert-Mayer-Str. 2, 60325 Frankfurt, Germany 
\and Laboratoire d’Astrophysique de Bordeaux, Universit\'e de Bordeaux, CNRS, B18N, 33615 Pessac, France
\and Instituto Nacional de Astrofísica, Óptica y Electrónica, Apartado Postal 51 y 216, 72000 Puebla, Mexico
}
        
\date{draft of \today}
\titlerunning{A proplyd-like object in Cygnus}  
\authorrunning{N. Schneider}  

\abstract{ A proplyd is defined as a young stellar object (YSO)
  surrounded by a circumstellar disk of gas and dust and embedded in a
  molecular envelope undergoing photo-evaporation by external
  ultraviolet (UV) radiation.  Since the discovery of the Orion
  proplyds, one question has arisen as to how inside-out
  photo-evaporation and external irradiation can influence the
  evolution of these systems. For such an investigation, it is
  essential to determine the molecular and atomic gas masses, as well
  as the photo-evaporation and free-fall timescales.  Understanding
  the dynamics within the photo-dissociation regions (PDRs) of a
  potential envelope–disc system, as well as the surrounding gas in
  relation to photo-evaporative flows, requires spectrally resolved
  line observations. Thus, we chose to investigate an isolated,
  globule-shaped object ($\sim0.37\, {\rm pc} \times 0.11\, {\rm pc}$
  at a distance of 1.4 kpc), located near the centre of the Cygnus OB2
  cluster and named proplyd \#7 in optical observations. In the
  literature, there is no consensus on the nature of this
  source. Observations point toward a massive star (with or without
  disc) with a \HII\ region or a G-type T Tauri star with a
  photo-evaporating disc, embedded in a molecular envelope.  We
  obtained a map of the \OI\ line at 63 $\mu$m with 6$''$ angular
  resolution and employed archival data of the \CII\ 158 $\mu$m line
  (14$''$ resolution), using the upGREAT heterodyne receiver aboard
  SOFIA. We also collected IRAM 30m CO data at 1mm (11$''$
  resolution). All the lines were detected across the whole object.
  The peak integrated \OI\ emission of $\sim$5 K km s$^{-1}$ is
  located $\sim$10$''$ west of an embedded YSO.  The \OI\ and
  \CII\ data near the source show bulk emission at $\sim$11 km
  s$^{-1}$ and a line wing at $\sim$13 km s$^{-1}$, while the
  $^{12}$CO 2$\to$1 data reveal additional blue-shifted high-velocity
  emission.  The widespread \OI\ emission prompts the question of its
  origin since the \OI\ line can serve as a cooling line for a PDR or
  for shocks associated with a disc.  From both local and non-local
  thermodynamic equilibrium (LTE and non-LTE) calculations, we
  obtained a column density of N$_{\rm OI}\approx$10$^{18}$ cm$^{-2}$
  at a density of 4-8 10$^3$~cm$^{-3}$.  The \OI\ line is, thus,
  sub-thermally excited. The KOSMA-$\tau$ PDR model can explain the
  emissions in the tail with a low external UV field ($<$350
  G$_\circ$, mostly consistent with our UV field estimates), but not
  at the location of the YSO. There, the high line intensities and
  increased line widths for all lines and a possible bipolar CO
  outflow suggest the presence of a protostellar disc. However, the
  existence of a thermal \HII\ region, revealed by combining existing
  and new radio continuum data, points towards a massive star -- and
  not a T Tauri-type one. The circumstellar environment of proplyd~\#7
  consists mostly of molecular gas. We derived molecular and atomic
  gas masses of $\sim$20 M$_\odot$ and a few M$_\odot$,
  respectively. The photo-evaporation (only considering external
  illumination) lifetime of 1.6$\times$10$^5$ yrs is shorter than the
  free-fall lifetime of 5.2$\times$10$^5$ yrs; thus, we find that
  proplyd~\#7 might not have had the time to produce many more
  stars. This viewpoint is supported by simulation results from the
  literature.  }
                
\keywords{interstellar medium: clouds
          -- individual objects: Cygnus X   
          -- molecules
          -- kinematics and dynamics
          -- Radio lines: ISM}
\maketitle

\section{Introduction} \label{sec:intro} 

\begin{figure*}
\centering \includegraphics[width=18.5cm,angle=0]{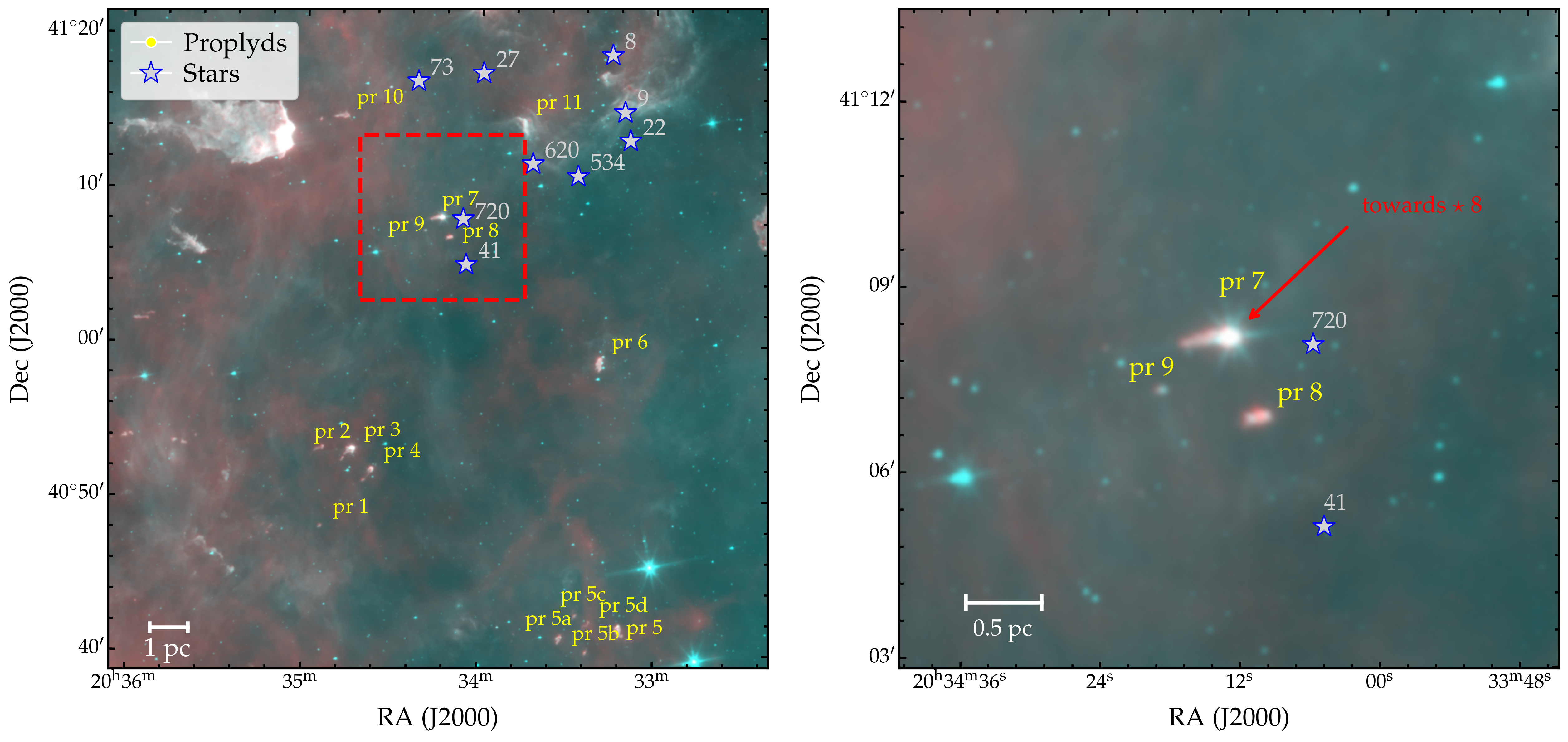}
\caption{ {False-colour image of the central Cygnus X region
    \citep[left;][]{Schneider2016a} and a zoom onto the region
    indicated by the red dashed box around Proplyd \#7 (right). The R
    and G channels correspond to Spitzer 8 $\mu$m, while the B channel
    corresponds to \textit{Herschel}} (PACS 70 $\mu$m) emission. The
  proplyd-like objects detected by \citet{Wright2012} and
  \citet{Schneider2016a} are labelled in yellow ({\tt
    pr1-11}). Proplyd \#7 ({\tt pr~7}) is located in the upper centre
  region, close to some massive stars of the Cyg OB2
  association. These are marked in grey, following the nomenclature in
  \citet{Wright2015}.  }
\label{overview}
\end{figure*}

Single low- and high-mass stars form through the gravitational
collapse of magnetised cores embedded within surrounding material. The
core's initial angular momentum leads to the development of a
Keplerian disc. While most of the mass accretes onto the star via the
envelope and disc, a small fraction forms planets. In the earliest
evolutionary stages, the material surrounding the star is
gas-rich. This `protoplanetary disc' is eventually dispersed by
internal (central star) and external (nearby massive stars)
photo-evaporation, by extreme-UV (EUV; energy $>$13.6 eV) and far-UV
(FUV; energy between 6 and 13.6 eV) photons and X-rays
\citep{Clarke2001,Ercolano2008,Gorti2008,Alexander2014,Winter2019,
  Winter2022,Allen2025}.  The EUV radiation dissociates molecules and
ionises atoms in the disc's outer layers while FUV radiation mostly
removes gas from the outer disc, where most of the mass is located
\citep{Gorti2009}. Atomic fine structure lines, such as the neutral
oxygen \OI\ 63 $\mu$m line, are observed in a layer of approximately
one A$_{\rm v}$ within the photodissociation region (PDR) on the disc
surface \citep{Dent2013,Gorti2008}.  The \OI\ 63 $\mu$m line is
predicted to be the strongest cooling line in discs, with line
luminosities as high as 10$^{-4}$ L$_\odot$ from T Tauri systems
\citep{Gorti2008,Meijerink2008,Aresu2014}.

The external photo-evaporation of the disc and its enveloping
molecular clump, caused by EUV and FUV radiation from nearby massive
stars, was first seen in optical observations of the Orion Nebula
\citep{Laques1979,Odell1993}. The discovered objects (named proplyds)
appear in silhouette as discs, surrounded by teardrop-shaped envelopes
with bright, ionised rims facing the exciting external source(s) and
tails extending away. Follow-up observations in other regions have
revealed considerable size variability among proplyds
\citep{Bally1998}, with a number of them shown to be irradiated
protoplanetary discs that are also still embedded in a molecular
envelope; thus, such objects could feasibly be considered
`globules'. Typical sizes for the discs of the Orion proplyds range
from 45 to 355 AU, with tails that can extend up to $\sim$2000 AU
\citep{Bally2000}.  Larger objects (i.e. with the size referring to
the longest extent of the globule-shaped disc+envelope structure) have
been identified, for example, in NGC 3603 (1.5-2.1$\times$10$^4$ AU,
\citealt{Brandner2000}) and Cygnus (5-10$\times$10$^4$ AU,
\citealt{Wright2012}). In Cygnus, it remains unclear if all such
objects possess discs, leading Wright et al. to describe them as
`proplyd-like'.  Proplyds have also been analytically and numerically
modelled \citep{Johnstone1998, Henney1999, Stoerzer1999, Richling2000,
  Carlsten2018, Peake2025}.  To distinguish true proplyds from other
UV-illuminated gas condensations without discs, such terms as
evaporating gaseous globules (EGG; \citealt{Smith2003}) and
globulettes \citep{Gahm2007} have been introduced for features around
10$^4$ AU ($\sim$0.05 pc) in size. \citet{McCaughrean2002} proposed
that either EGGs are a precursor to proplyds that eventually go on to
form stars or the gas ends up completely photo-evaporating.  In
\citet{Schneider2016a}, EGGs, globules, and proplyd-like objects in
Cygnus X were classified using {\textit{Herschel}} far-infrared (FIR)
imaging at 70 $\mu$m.

Studying these objects is of particular interest, as they may
represent an isolated mode of star formation, possibly regulated by
gas compression driven by radiation or stellar winds. In such cases,
the external compression can lead to a shortened star formation
timescale. These objects are convenient to study because they exhibit
a well-defined morphology and are located outside major molecular
cloud complexes. Observational studies, such as the present
investigation of a single object in Cygnus X (proplyd~\#7), together
with numerical simulations, are essential for probing the interaction
between protoplanetary systems and their surrounding
environments. This paper is aimed at improving our understanding of
the physical nature of this specific object and to contribute to our
ongoing multi-wavelength studies
\citep{Schneider2016a,Djupvik2017,Schneider2021} of features such as
globules, pillars, EGGs, and proplyds that have been shaped by stellar
feedback.

Here, we present observations of spectrally resolved \OI\ 63~$\mu$m
and ionised carbon \CII\ 158~$\mu$m line emission from proplyd \#7 at
an angular resolution of 6$''$ and 14$''$, respectively, together with
complementary molecular line observations at 1mm
(Sect.~\ref{sec:obs}).  For simplicity, we call the whole feature, the
possible disc, and the enveloping gas including the tail and proplyd
(or, specifically, proplyd \#7). We assume a distance of 1.4 kpc for
proplyd \#7 since it is most likely shaped by the Cyg OB2 association
which is located at that distance \citep{Rygl2012,Dzib2013}.  Previous
studies, particularly Gas Survey of Protoplanetary Systems (GASPS;
\citealt{Dent2013}) and recent investigations of proplyds in Orion and
Carina \citep{Champion2017}, have relied on velocity-unresolved data
obtained from the PACS spectrometer on board {\textit{Herschel}} with
a resolution of approximately 9$''$. By utilising the enhanced
velocity resolution provided by upGREAT on the Stratospheric
Observatory for Far-Infrared Astronomy (SOFIA), we can more
effectively characterise the gas dynamics associated with the
proplyd's envelope (Sect.~\ref{sec:results}). However, considering the
angular resolution of 6$''$ (corresponding to $\sim$8400 AU at a
distance of 1.4 kpc), it is not straightforward to resolve the
transition between the (potential) protoplanetary disc and the
surrounding envelope. We employed the FIR line and continuum data,
along with the CO 2$\to$1 observations at two positions of proplyd
\#7, for the PDR modeling and to independently determine the physical
properties such as column densities and masses, as well as the radio
SED (Sect. \ref{sec:analysis}). Section~\ref{sec:discuss} addresses
the lifetime of this object and provides interpretations of its
nature.

\section{Proplyd \#7 in Cygnus X} \label{sec:proplyd}

This study focuses on an object in Cygnus X classified in the infrared
(IR) as IRAS 20324+4057 and by \citet{Wright2012} as a proplyd-like
feature. Figure~\ref{overview} shows these objects that appear in
projection close to the Cyg OB2 cluster centre
\citep{Schneider2016a}. The source we study here (indicated in
Fig.~\ref{overview} as 'pr~7') was previously investigated by several
authors; however, its true nature remains uncertain.

\citet{Comeron2002} classified the source as a potential OB star
(designated star B2), based on near-IR (NIR) spectroscopy, in
particular the Br-$\gamma$ line.  \citet{Pereira2007} identified it as
an \HII\ region (source GLMP1000) with a bow shock or a disc with
outflow, deduced from the observed H$\alpha$/\NII\ and
H$\alpha$/\SII\ line intensity ratios. The ratios are more compatible
with a photoionised region than a shock-excited nebulosity. H$\alpha$
and \NII\ images (their Fig. B.1) show that the north-western edge of
proplyd \#7 is particularly bright and conclude that B2 could be the
(internal) exciting source. \citet{Wright2012} combined H$\alpha$,
\textit{Hubble} Space Telescope (HST), and \textit{Spitzer}
observations, named the object proplyd \#7 (a term adopted in this
paper) exhibiting a typical head-tail structure with a rim-brightened
head. \citet{Sahai2012} further investigated this source, named it
'Tadpole', and proposed the term frEGG (free-floating EGG). They
determined that it has a length of 54.7$''$ (7.7$\times$10$^4$ AU,
0.37 pc) and a maximum width of 14.1$''$ (2.0$\times$10$^4$ AU, 0.1
pc). The size of proplyd \#7 is thus at the upper end of the size
range mentioned above. They identified a `cometary knot' in HST images
(their Fig. 2), with a peak at RA(2000) = 20$^h$34$^m$13.23$^s$,
Dec(2000) = 41$^\circ$08$'$14.6$''$, coinciding with the location of a
luminous point source in the 2MASS (20341326+4108140) and MSX6c
(G080.1909+00.5353) catalogues, and with star B2 in
\citet{Comeron2002}. It is interpreted as scattered light from the
lobes of a collimated bipolar outflow directed along the tilted, polar
axis of a flared disc (or dense equatorial region). In addition, they
detected three faint red stars north-west from the knot, using the
High Angular Resolution Observer NIR camera on the Palomar 200-inch
telescope.  The VLA map of the radio emission at 8.5 GHz (Fig. 3 in
\citet{Sahai2012} strongly peaks at the head of proplyd \#7. The
authors suggest that this is a result of a compressed magnetic layer
in this front that is interacting with cosmic rays (CRs) associated
with the Cyg OB2 association. From CO observations (45$''$ beam), they
derived a mass of 3.7 M$_\odot$ of the molecular gas around the
central object; thus, they consider it more plausible that proplyd \#7
is an EGG. \citet{Guarcello2013} classified the central young stellar
object (YSO) in proplyd~\#7 as a Class I star (star 1724) from the
OSIRIS (Optical System for Imaging and low-Resolution Integrated
Spectroscopy) Gran Telescopio Canaria survey. In a subsequent study by
\citet{Guarcello2014}, the authors tentatively propose that the YSO's
spectral features are consistent with a young ($<$2 Myr) G-type star
(T Tauri), actively accreting from a disc.  They reported an intense
outflow and a potential stellar jet of ionised gas, and they detected
a brightness variability. They found that the mass loss rate is higher
than the accretion rate and suggested that the outflow
dominates. \citet{Isequilla2019} deduced from their observations at
325 and 610 MHz, using the Giant Meterwave Radio Telescope (GMRT) that
the radio emission from proplyd \#7 is of thermal origin, in contrast
to what is stated in \citet{Sahai2012}.

From the {\textit{Herschel}} data at 36$''$ angular resolution,
\citet{Schneider2016a} derived a hydrogen density of $8.8 \times 10^3$
cm$^{-3}$, a dust temperature of 18.3 K, a size of 0.37 $\times$ 0.11
pc (same as derived by \citealt{Sahai2012}), and a mass of 43
M$_\odot$ for proplyd \#7. These values refer to the disc+envelope
system and the tail. Proplyd \#7 is located at 5.9 pc projected
distance from the Cyg OB2 association centre, commonly defined by the
Cyg OB2 \#8 \citep{Schulte1958,Wright2015} trapezium of O stars. It
has been suggested \citep{Wright2012,Guarcello2014} that the
morphology and orientation of proplyd \#7 (and other objects in this
region) is governed by the stellar feedback from Cyg OB2 \#8. However,
\citet{Isequilla2019} favour the O-stars Cyg~OB2 \#9 and \#22 as the
dominating ionizing sources. The closest system to proplyd \#7 are
three B-stars of type B0.5V, B1.5V, and B1V. We derive and discuss the
UV field in this paper.

\begin{figure*}
\centering
\includegraphics[width=18cm]{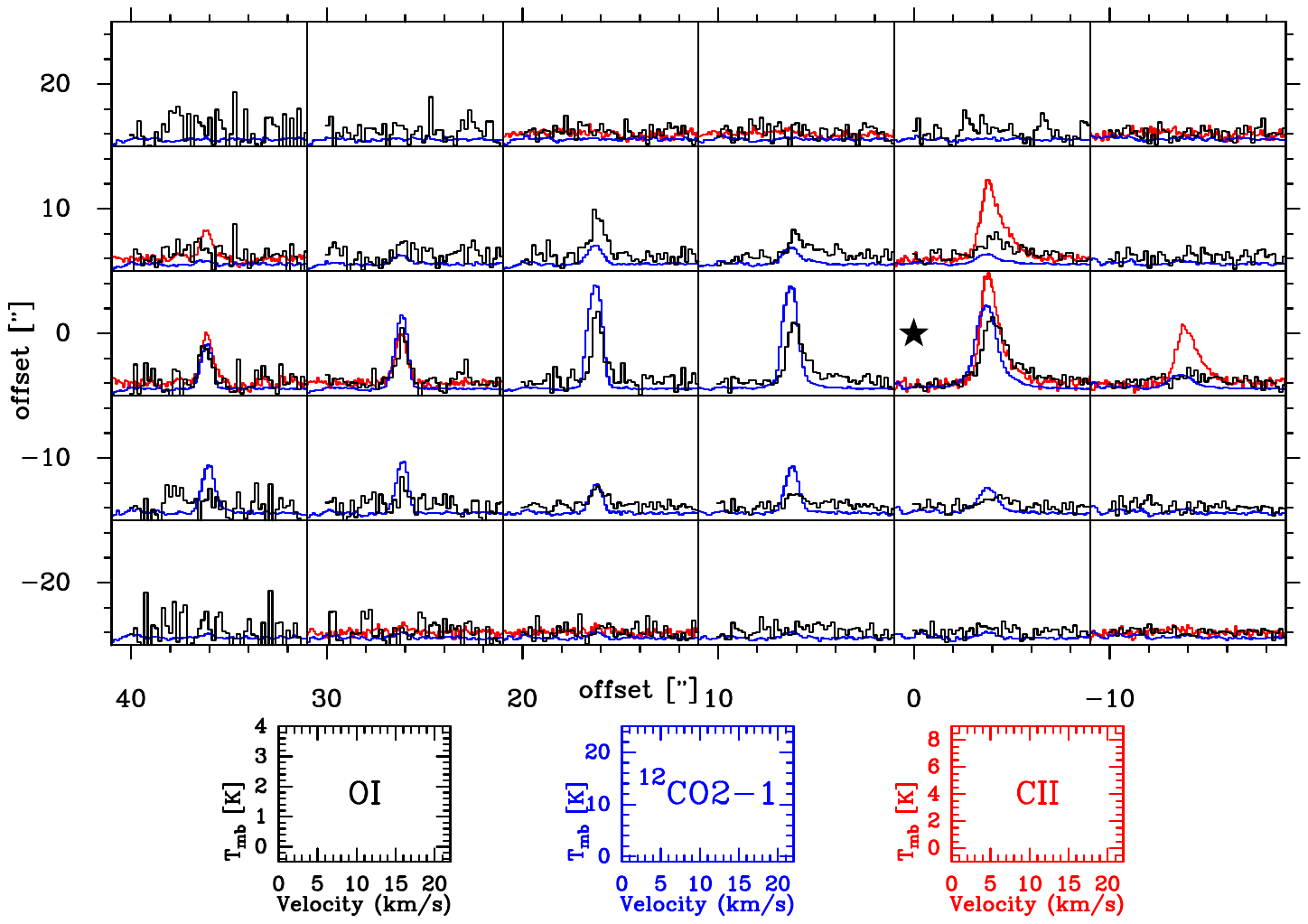}
\caption{ \OI\ 63 $\mu$m, \CII\ 158 $\mu$m, and $^{12}$CO 2$\to$1
  spectra of proplyd \#7. To increase the S/N, the \OI\ data were
  smoothed to an angular resolution of 10$''$ and re-gridded to 10$''$
  (original resolution 6$''$ on a $\sim$2$''$ sampling grid). The
  velocity resolution was degraded to $\sim$0.4 km s$^{-1}$. The
  \CII\ (CO) spectra are on their nominal angular resolution of 14$''$
  (12$''$) and velocity resolution of 0.2 (0.3) km s$^{-1}$. The
  displayed main beam brightness temperature and velocity ranges are
  indicated in the small lower panels. The star indicates the
  approximate location of the YSO.  }
\label{spectra}
\end{figure*}

\begin{figure}
\centering
\includegraphics[width=8cm]{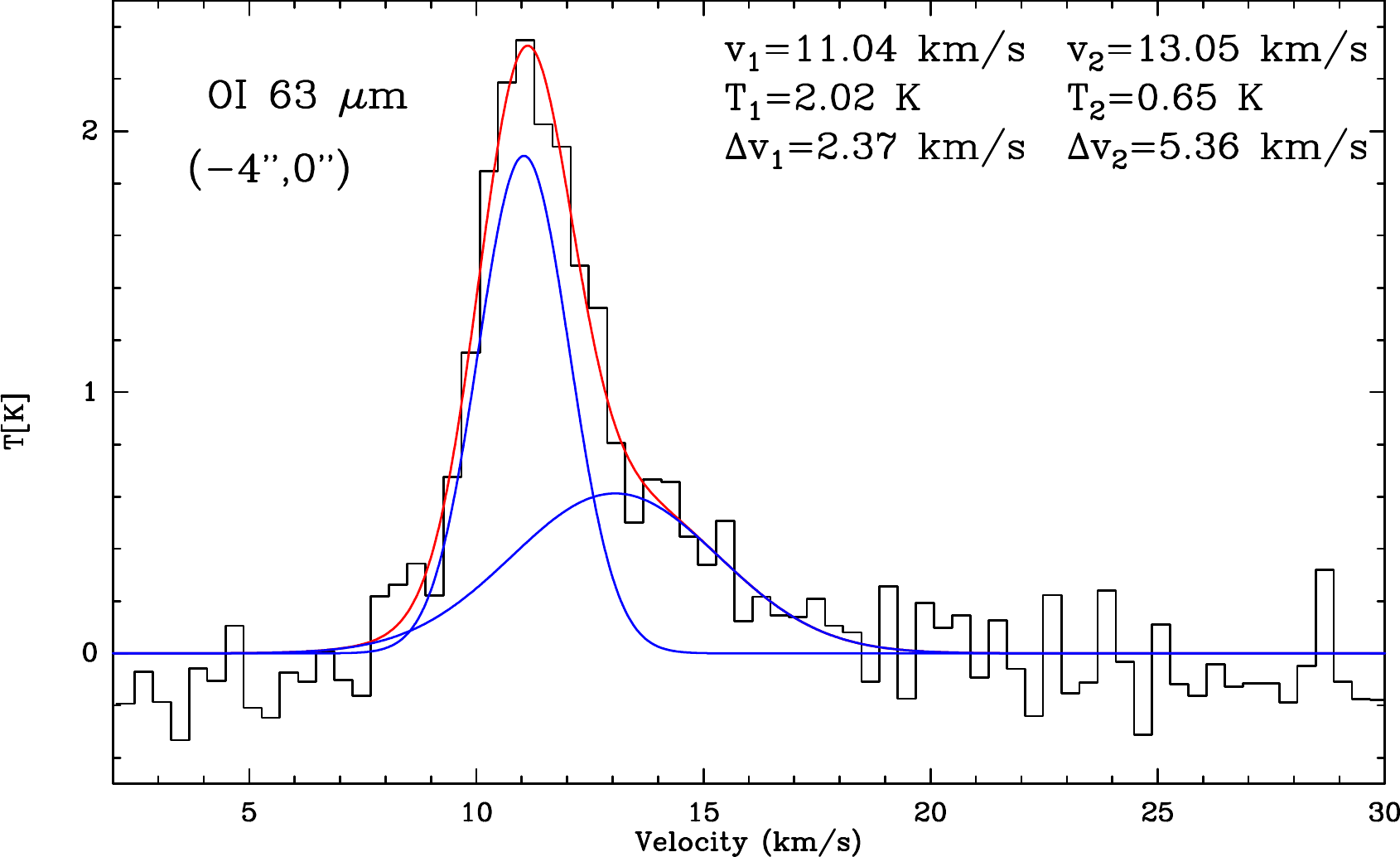}
\caption{ Example of an \OI\ spectrum and fit to the data. The black
  histogram shows the observed \OI\ spectrum closest to the central
  YSO (position with offset $-$4$''$,0), smoothed to an angular
  resolution of 10$''$. The blue Gaussian lines represent the fit to
  this spectrum with two components.  Centre velocity, temperature,
  and FWHM values are given in the panel. The red line gives the
  resulting fit.  }
\label{spectra-fit}
\end{figure}

\begin{figure}[htbp]
\begin{center} 
\includegraphics [width=8.5cm, angle={0}]{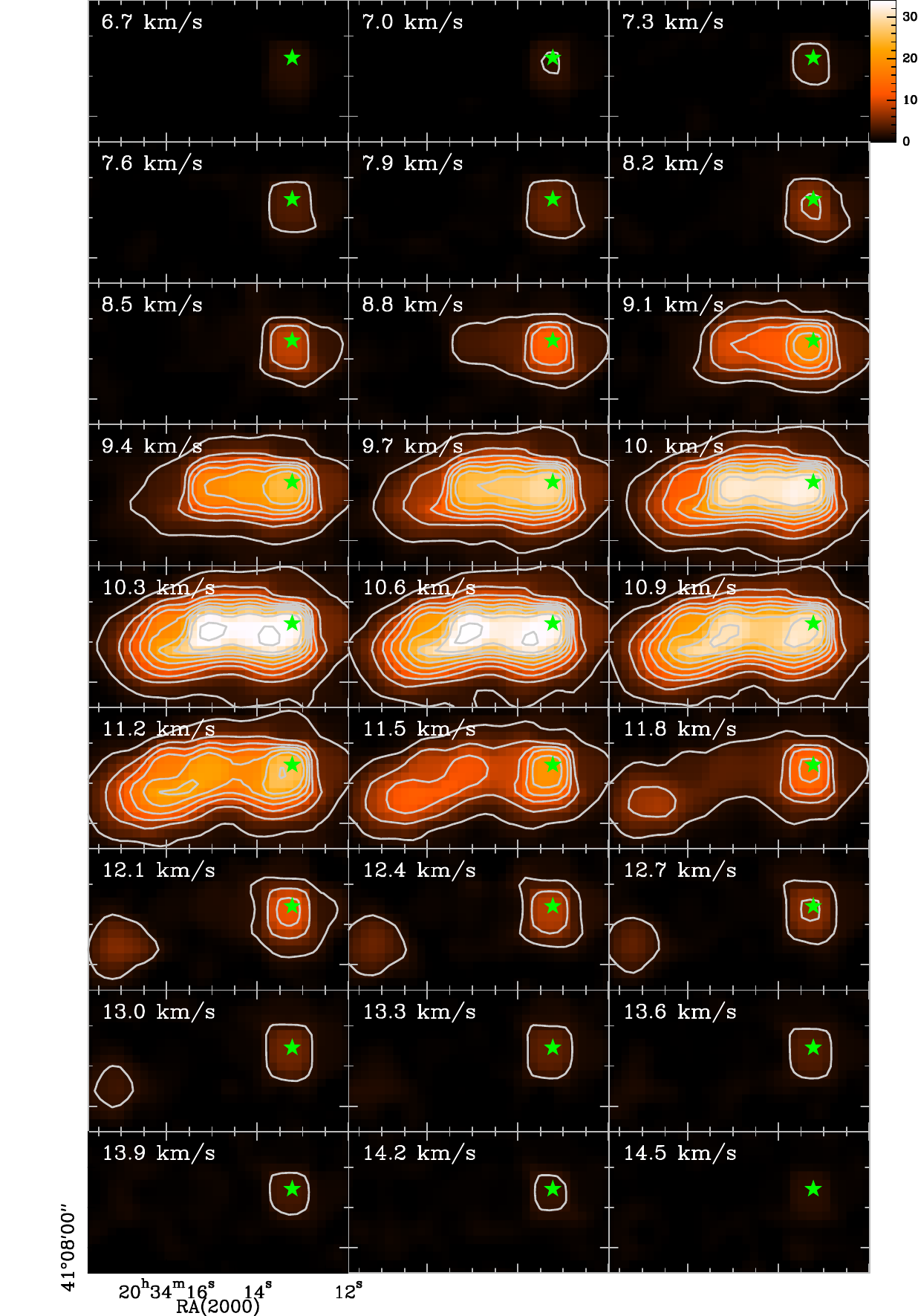} 
\caption{ Channel map of $^{12}$CO 2$\to$1 line emission. Each channel
  covers a velocity range of 0.3 km s$^{-1}$. Contours go from 2 to 34
  K km s$^{-1}$ in steps of 4 K km s$^{-1}$ and the green star
  indicates the position of the YSO.  }
\label{fig:co-channels}
\end{center} 
\end{figure}

\section{Observations} \label{sec:obs} 
\subsection{SOFIA} \label{sec:sofia} 

The \OI\ atomic fine structure line at 4.74478 THz (63.2 $\mu$m) and the CO 16$\to$15 rotational line at 1.841345 THz were observed with the heterodyne receiver upGREAT
\footnote{German Receiver for Astronomy at Terahertz. GREAT is a
development by the MPI f\"ur Radioastronomie and the
KOSMA/Universit\"at zu K\"oln, in cooperation with the MPI f\"ur
Sonnensystemforschung and the DLR Institut f\"ur Planetenforschung.}
on board SOFIA during one guaranteed time (GT) flight on 2 November
2016, from Palmdale, California (programme ID 83\_0433). The \OI\ line
was observed in the 7-pixel High Frequency Array (HFA), and the CO
line in the one-pixel L2 channel.  The map centre position is
RA(2000)=20$^h$34$^m$13.3$^s$ and
Dec(2000)=+41$^\circ$08$'$13.8$''$. The observing mode was set to
on-the-fly (OTF) mapping with chopping in single phase A with a chop
throw of 120$''$ and a chop frequency of 0.655 Hz. The array
orientation was tilted by --19.1 deg relative to the scanning
direction so that the HFA 7 pixels scan at an equal distance. The
central channel was set to a velocity of +8 km s$^{-1}$. The map
covers an area of $\sim$60$'' \times$30$''$.

\begin{figure*}
\centering
\includegraphics[width=9cm]{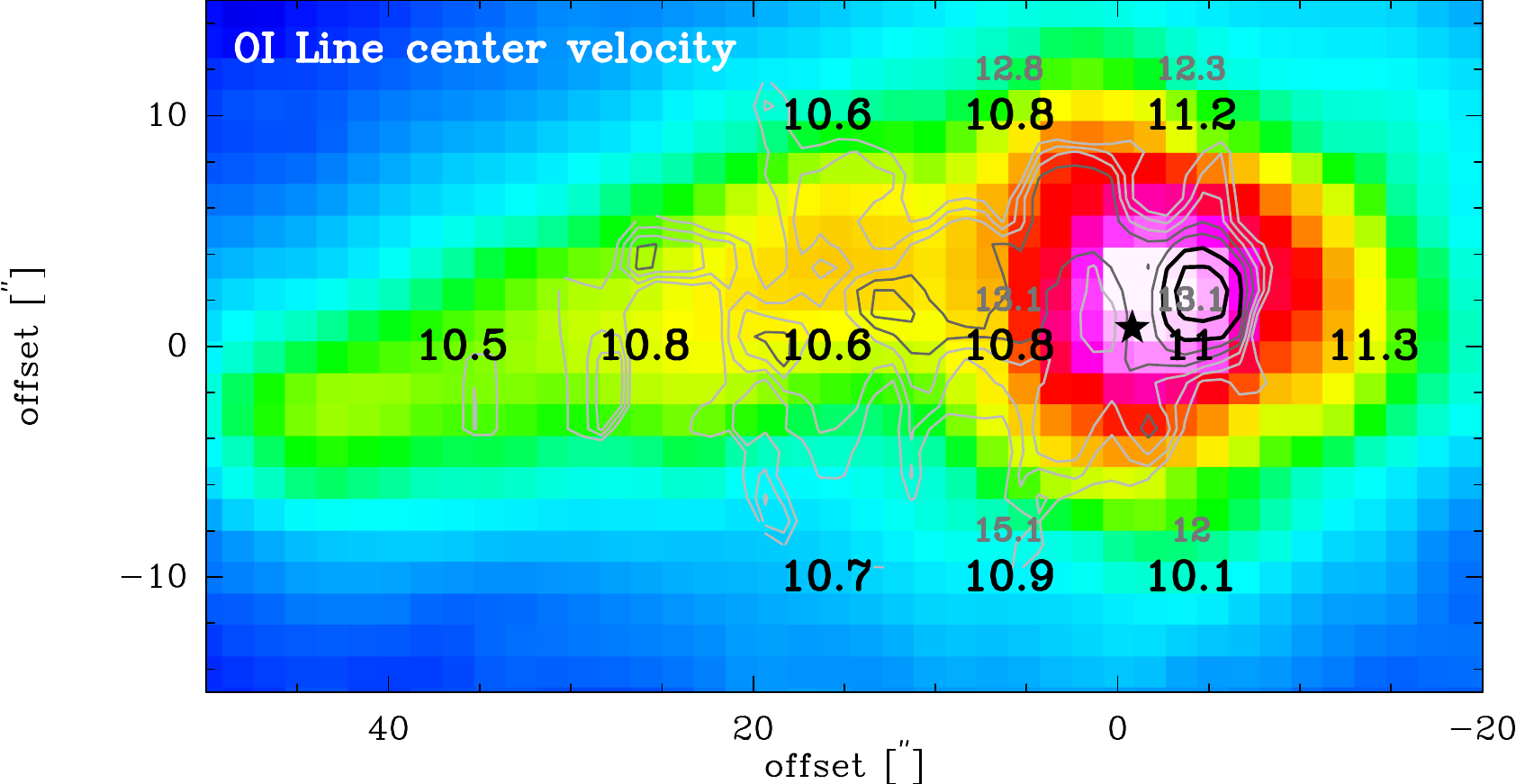}
\includegraphics[width=9cm]{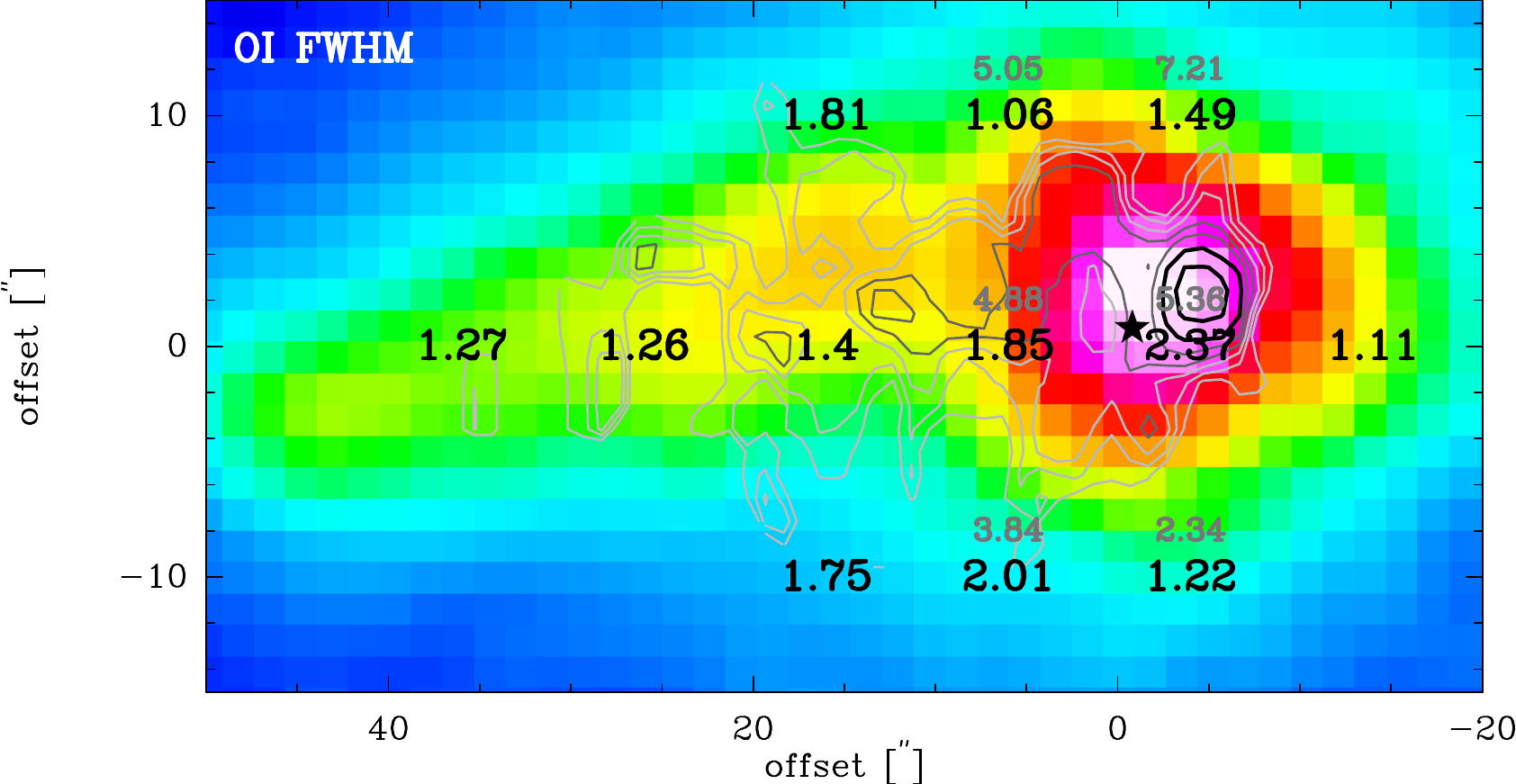}
\caption{ Distribution of \OI\ line centre velocity and FWHM. Overlays
  of \OI\ line integrated (v=9 to 15 km s$^{-1}$) emission (contour
  levels 5, 6, 7, 8, and 8.5 K km s$^{-1}$) on
  {\textit{Herschel}}/PACS 70 $\mu$m emission in color.  The left
  panel shows the line centre velocities in km s$^{-1}$ of the two
  components (first component in black, second in grey) and the right
  panel the linewidth in km s$^{-1}$. The values (Table~\ref{tab1})
  were determined from Gaussian fits to the spectra from
  Fig.~\ref{spectra}.  }
\label{fit-values}
\end{figure*}

\begin{figure}
\centering
\includegraphics[width=9cm]{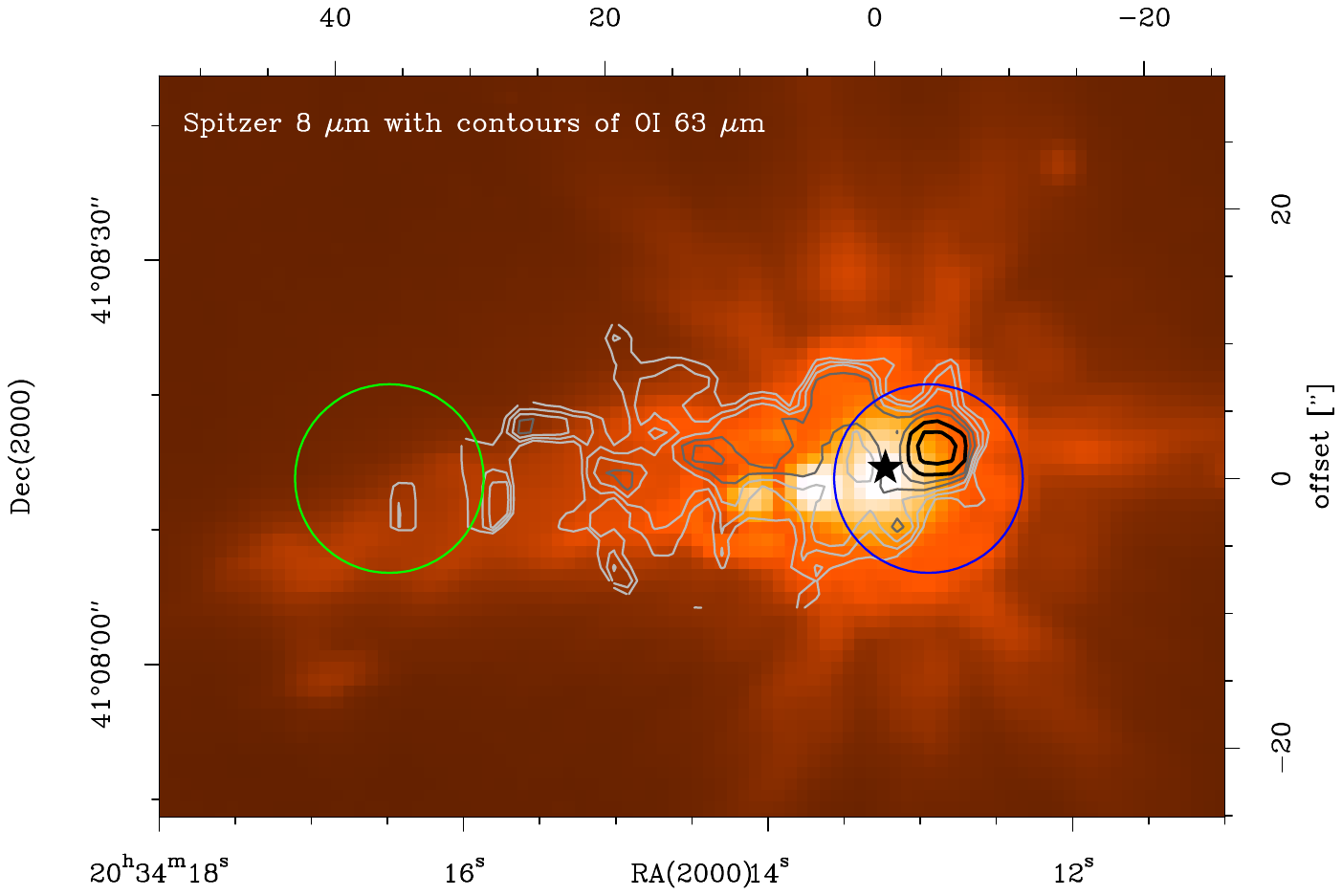}
\caption{ \textit{Spitzer} 8 $\mu$m image of proplyd \#7. The NIR map
  is in units of MJy/sr, ranging from 44 to 1.4$\times$10$^4$ with
  contours of \OI\ emission (5, 6, 7, 8, 8.5 K km s$^{-1}$)
  overlaid. Two positions where the PDR modelling was performed are
  indicated: position 1 (offset $-$4$''$,0) with a blue circle (radius
  7$''$) and position 2 (offset 36$''$,0) with a green circle,
  respectively.  }
\label{irac}
\end{figure}

Procedures to determine the instrument alignment and telescope
efficiencies, antenna temperature, and atmospheric calibration, as
well as the spectrometers used are described in \citet{Heyminck2012}
and \citet{Guan2012}. All line intensities are reported as main beam
temperatures scaled with main-beam efficiencies of 0.69 and 0.68 for
\OI\ and CO, respectively, and a forward efficiency of 0.97. The main
beam sizes are 6.1$''$ for the H-channel and 15.3$''$ for the L2
channel.

The calibrated \OI\ and CO spectra were further reduced and analysed
with the GILDAS software, developed and maintained by IRAM. From the
spectra, a third-order baseline was removed and spectra were then
averaged with 1/$\sigma^2$ weighting (baseline noise).  The mean rms
(root mean square) noise temperatures per 0.28 km s$^{-1}$ velocity
bin is 0.8 K for \OI\ in a 6$''$ beam. The CO 16$\to$15 line was not
detected on a 0.4 K rms level in a velocity bin of 0.6 km s$^{-1}$. We
estimate that the absolute calibration uncertainty is around
$\sim$20\%.  The telluric \OI\ line, originating from the mesosphere,
is visible in the spectra as an absorption feature, but is located at
+1 km s$^{-1}$ and, thus, far away from the systemic velocity of the
proplyd, which is around 11 km s$^{-1}$.

We also used \OI\ 63 $\mu$m and \CII\ 158 $\mu$m line data from the
SOFIA archive, collected as part of programme 05\_0176 (PIs: R. Sahai
et al.), obtained with the HFA and LFA-array of upGREAT
\citep{Risacher2018}. The spectra had a baseline of an order of one
removed in terms of the fixed velocity scale and they were also
sigma-weighted during the averaging of spectra per position.  The beam
efficiencies were determined for all HFA (LFA) pixels (varying between
0.59 and 0.68 for the HFA array) and between 0.65 and 0.73 for the LFA
array using Jupiter as a calibrator and applied individually. All
\OI\ and \CII\ temperatures are thus in main beam brightness
temperatures and the beam size is 14.1$''$ for \CII\ and 6.1$''$ for
\OI. We note that only a single footprint of the arrays was obtained
for proplyd \#7 and not a full map in \OI\ as for our GT data set. A
complementary set of publicly available data were used for this study,
including {\textit{Herschel}} and \textit{Spitzer}.

\subsection{IRAM 30m} \label{sec:iram} 
Proplyd \#7 was observed for 2h within Director’s Discretionary Time
(DTT) in a setting at 1mm (tuning frequency 218.6 GHz) with EMIR,
covering the following lines: $^{12}$CO (230.538 GHz), $^{13}$CO
(220.3987 GHz), and C$^{18}$O 2$\to$1 (219.560 GHz), as well as other
molecular lines such as SiO 5$\to$4, DCO$^+$ 3$\to$2, and DCN 3$\to$2,
and two H$_2$CO lines. All data have an angular resolution of
11-12$''$. Here, we present only the isotopomeric CO 2$\to$1 data. We
started by observing a strip with four pointings and then performed an
OTF map that is 120$''\times$80$''$ in size, with one horizontal and
one vertical coverage, respectively. The weather conditions were good
(opacity at 1mm 0.3 to 0.4) and the pointing accuracy was checked with
the source K3-50A and found to be good within 1-2$''$.

\subsection{GLOSTAR} \label{sec:glostar-obs} 
We made use of radio continuum data from the Global View of Star
Formation in the Milky Way (GLOSTAR) survey
\citep{Medina2019,Brunthaler2021}, which combines data taken at the
Karl G. Jansky Very Large Array (VLA) in its D and B-configurations
with zero-spacing from the Effelsberg 100m telescope. The correlator
setting covers the continuum emission in full polarization within 4–8
GHz (using two 1 GHz sub-bands centred at 4.7 and 6.9 GHz).  For
technical details, we refer to the papers cited above. Here, we used
data in a combined B+D configuration with a resolution of 4$''$. The
mean background level, as determined in an emission-free region
north-west of proplyd ~\#7, is 0.2~mJy.

\begin{table*} 
  \caption{Results  from fitting two Gaussian lines to the observed \OI\ spectra. }
\label{tab1}
\centering  
\begin{tabular}{c|lllll|llll}     
\hline\hline
         &      &            &      &             &                 &                &       &             &             \\     
Offset   & rms  & I(int)$_1$   & T$_1$& v$_1$       & $\Delta{\rm v}_1$& I(int)$_2$      & T$_2$ & v$_2$       &  $\Delta{\rm v}_2$   \\ 
($''$,$''$)& [K]&[K kms$^{-1}$]& [K]&[kms$^{-1}$] & [kms$^{-1}$]    &[K kms$^{-1}$]  & [K]   & [kms$^{-1}$]& [kms$^{-1}$]  \\     
\hline 
 -4, -10 & 0.10 & 0.21$\pm$0.19 & 0.17 & 10.14$\pm$0.40 & 1.22$\pm$0.90& 0.91$\pm$0.23 & 0.39 & 12.00$\pm$0.40 & 2.34$\pm$0.52 \\ 
  6, -10 & 0.12 & 1.01$\pm$0.20 & 0.50 & 10.91$\pm$0.21 & 2.01$\pm$0.46& 0.61$\pm$0.28& 0.16 & 15.15$\pm$0.88 & 3.84$\pm$2.18 \\
 16, -10 & 0.11 & 1.15$\pm$0.25 & 0.66 & 10.74$\pm$0.11 & 1.75$\pm$0.32&     -        &  -   &  -            & - \\
-14, 0   & 0.14 & 0.38$\pm$0.21 & 0.30 & 11.28$\pm$0.29 & 1.11$\pm$0.72&     -        &  -    &  -            & - \\
 -4, 0   & 0.14 & 4.81$\pm$0.34 & 2.02 & 11.04$\pm$0.07 & 2.37$\pm$0.14& 3.50$\pm$0.57& 0.65 & 13.05$\pm$0.19 & 5.37$\pm$0.84 \\
  6, 0   & 0.16 & 3.86$\pm$0.53 & 2.09 & 10.76$\pm$0.07 & 1.85$\pm$0.18& 1.85$\pm$0.56& 0.37 & 13.12$\pm$0.90 & 4.88$\pm$1.34 \\
 16, 0   & 0.24 & 3.54$\pm$0.22 & 2.52 & 10.60$\pm$0.04 & 1.41$\pm$0.07&     -        &  -   &  -             & - \\ 
 26, 0   & 0.27 & 2.35$\pm$0.30 & 1.88 & 10.78$\pm$0.07 & 1.26$\pm$0.21&     -        &  -   &  -             & - \\
 36, 0   & 0.30 & 2.00$\pm$0.35 & 1.57 & 10.49$\pm$0.13 & 1.27$\pm$0.16&     -        &  -   &  -             & - \\
 -4, 10  & 0.15 & 0.81$\pm$0.45 & 0.54 & 11.23$\pm$0.19 & 1.49$\pm$0.72&2.56$\pm$0.59 & 0.36 &  12.26$\pm$0.82 & 7.21$\pm$1.46 \\
  6, 10  & 0.13 & 0.85$\pm$0.22 & 0.81 & 10.83$\pm$0.08 & 1.06$\pm$0.25&1.87$\pm$0.36 & 0.37 &  12.76$\pm$0.55 & 5.05$\pm$0.97 \\
 16, 10  & 0.19 & 2.88$\pm$0.17 & 1.60 & 10.65$\pm$0.40 & 1.81$\pm$0.40&     -        &  -   &  -            &  \\
\end{tabular}
\tablefoot{The rms in the second column refers to a channel width of
  0.4 km s$^{-1}$. The other parameters are line integrated intensity,
  I(int), peak temperature, T, centre velocity, v, and line width,
  $\Delta{\rm v}$. The stronger line is at lower velocities and has an
  index of 1. The weaker higher velocity component with an index of 2
  cannot be fitted for all positions. The results of the centre line
  velocities and line widths are displayed in Fig~\ref{fit-values}. }
\end{table*}

\section{Results} \label{sec:results}

\subsection{[OI], [CII], and CO line observations} \label{subsec:lines}
Figure~\ref{spectra} presents the observed \OI, \CII, and CO 2$\to$1
spectra across proplyd \#7. The \OI\ line is prominently detected,
with peak emission around the (0,0) position, very close to the
location of the YSO. Notably, strong \OI\ emission is also observed in
the proplyd tail (1.6–1.9 K). However, to thermally excite bright
\OI\ 63~$\mu$m emission densities of a few 10$^5$ cm$^{-3}$ and high
temperatures (typically $>$100 K) are required. As we will see later,
these conditions are not given in the proplyd tail. The
\CII\ observations tend to feature fewer data points, but they do
exhibit a similar spatial distribution for the emission. In the
proplyd tail, all lines display a single Gaussian component with a
peak velocity around 10.5~km~s$^{-1}$. In contrast, the spectra around
the (0,0) position require a fit with at least two\footnote{We did not
try a fit to the blue-shifted component, that is visible in the CO
spectra and to a lesser extent in the \CII\ spectra, because of the
low S/N in the \OI\ data.}  Gaussian components (an additional
component at $\sim$13 km s$^{-1}$), an example for a fit to the
\OI\ line at offset ($-$4$''$,0) is displayed in
Fig.~\ref{spectra-fit}.  The red-shifted wing is seen in all lines and
most likely indicates gas streaming away from the observer.  Although
low-mass stars often exhibit outflow signatures during their early
evolutionary stages driven by accretion processes in the disc, with
previous studies, such as \citet{Guarcello2014}, indicating outflow
activity, it is unlikely that the 13 km s$^{-1}$ component represents
a highly collimated \OI\ jet. This is because typical \OI\ jets have
velocities exceeding 100 km s$^{-1}$ \citep{Podio2012}. However, for
massive stars \citep{Kuiper2011}, an envelope-disc interaction leads
to an atomic jet of at least partly ionised gas orthogonal to the
disc, which can then drive a molecular outflow (see below) and create
an outflow-confined \HII\ region. The \OI\ velocities can be lower,
but still of the order of a few 10 km s$^{-1}$ and thus also faster
than the velocity difference we observe for the wing emission with
respect to the bulk emission.

Another possibility is that the red-shifted \OI, \CII, and CO emission
is due to a photo-evaporative flow, triggered by external UV
radiation. Proplyd \#7 is clearly exposed to the radiation and wind of
the nearby O-stars (Sect.~\ref{subsec:fuv}) and we could expect a
dynamic impact with gas released from the surface layer. This would
explain the small difference in velocity between the bulk emission and
the red-shifted wing emission. However, the atomic and molecular line
profiles across proplyd \#7 do not show broad wings everywhere, which
would be expected in the case of significant photo-evaporation. We
note that this can be a geometry effect, combined with the low angular
resolution. In the case where the closest star \#720 (see
Sect.~\ref{subsec:fuv}) would have the largest influence on proplyd
\#7, the radiation field and wind would be strongest at the western
border, which is only approximately two beam sizes away from the
YSO. The tail extends further east from the YSO and is thus shielded
from direct stellar irradiation.

Table~\ref{tab1} presents the \OI\ line fitting results and associated
errors for both spectral components. Only positions where the main
beam brightness temperature exceeds the rms noise are included, as
indicated in the table. However, some results have substantial
uncertainties and should be interpreted with
caution. Figure~\ref{fit-values} displays maps of the line centre
velocity and width overlaid on a colour plot of PACS 70 $\mu$m
emission, with line-integrated (9 to 14 km s$^{-1}$) \OI\ emission
contours. The peak of \OI\ emission is offset by approximately 6$''$
to the west of the YSO and the 70 $\mu$m peak, which marks the
location of hot dust heated by the central star.\footnote{We note that
the dust temperature of proplyd \#7 is 18.3 K (Table 1 in
\citealt{Schneider2016a}), the highest values amongst all proplyd-like
objects in the Cygnus sample, which may point towards internal
excitation. However, this is a spatially averaged value of a map at
36$''$ resolution (Fig. 8 in \citealt{Schneider2016a}) and still
reflects mostly the cooler envelope gas.}  A similar distribution is
evident in the \textit{Spitzer}/IRAC image at 8 $\mu$m
(Fig.~\ref{irac}).

From Fig.~\ref{fit-values}, a clear gradient in line velocity and
width of the first component is observed, extending from east to west
along proplyd \#7. The central velocity decreases from 11.3 km
s$^{-1}$ at the western edge to 10.5 km s$^{-1}$ at the eastern end of
the tail. The line width of the main component 1 is approximately 1.1
km s$^{-1}$ near the western border, increases significantly at the
position of the YSO to 2.37 km s$^{-1}$ and gradually decreases to
1.27 km s$^{-1}$ in the tail. The elevated line width at the central
position, combined with the presence of a high-velocity wing in
\OI\ emission, suggests a significant role for the central star’s
radiation in producing the observed \OI\ emission (see discussion
above). If the \OI\ 63 $\mu$m line was mainly driven by the externally
illuminated PDR, such a marked variation in line width would not be
expected. If the central source is a massive star with a (compact)
\HII\ region, the interaction of the radiation and wind of the source
with the interface (i.e. the atomic PDR and then the molecular
envelope) could result in high-velocity emission in all line tracers
(\OI, and \CII\ in the PDR and CO in the molecular gas) without the
need of an accretion disc. If there is an accretion disc (regardless
if the central source is a T Tauri or a massive star), the high
accretion luminosity could be the source for the \OI\ excitation by
shocks though accretion luminosity is also often released in the UV
and then leads to a second internal PDR. This would superimpose the
shock effect. The \OI\ (and possible \CII) emission would then arise
from the photoevaporating disc. However, we did not detect
emission\footnote{The rms in a channel of 0.27 km s$^{-1}$ width was
only 0.3 K so that a weak SiO 5$\to$4 line cannot be excluded.} from a
shock tracer (SiO~5$\to$4) and the observed velocity gradient across
the proplyd is also not straightforward to explain in this scenario.

\subsection{A protostellar CO outflow} \label{subsec:outflow}
A CO 2$\to$1 bipolar outflow, oriented apparently mostly along the
line of sight, is clearly identifiable in the channel maps of CO
emission shown in Fig.~\ref{fig:co-channels} (channel maps of
$^{13}$CO and C$^{18}$O 2$\to$1 emission are given in
Fig.~\ref{fig:IRAM2}), and in the 3D position-velocity (PV) cut in
Fig.~\ref{fig:3D-PV}. Both, the blue-shifted ($\sim$7 to 9
km~s$^{-1}$) and the red-shifted ($\sim$11 to 14 km s$^{-1}$)
components appear rather focussed on the central source. One
possibility is that we are observing a protostellar outflow from a
young star with an accretion disc that lies mostly in the plane of the
sky. This is in tension with the proposal of \cite{Sahai2012}, who
interpreted a conical optical feature as scattered light from an
outflow, tilted in a way so that the near-side of the disc is lying to
the north-west of the YSO. They derived an opening angle of the
bipolar outflow cavity around 5-7$^\circ$ and an outflow cavity axis
inclined at an angle of 18$^\circ$ to the line of sight. Because of
the low angular resolution of our CO data, we could exclude this
geometry, but we do note that comparing the CO outflow and the optical
observations is challenging. Whether the central source is a low-mass
star, a G-type star (T Tauri) as proposed by \citet{Guarcello2014} or
a massive young star can also not be concluded from the CO
observations alone.

Another possibility for the outflow emission is that a massive star
(with or without disc) has created a cavity by its radiation and winds
and entrained gas can also arise from ablation of gas at the cavity
borders to the surrounding molecular cloud (see
Sect. \ref{subsec:lines}). This scenario was previously reported, for
example, in S106 \citep{Schneider2018} and in a globule in Cygnus X
\citep{Djupvik2017,Schneider2021}.

\begin{figure}[htbp]
\begin{center} 
\includegraphics [width=9cm, angle={0}]{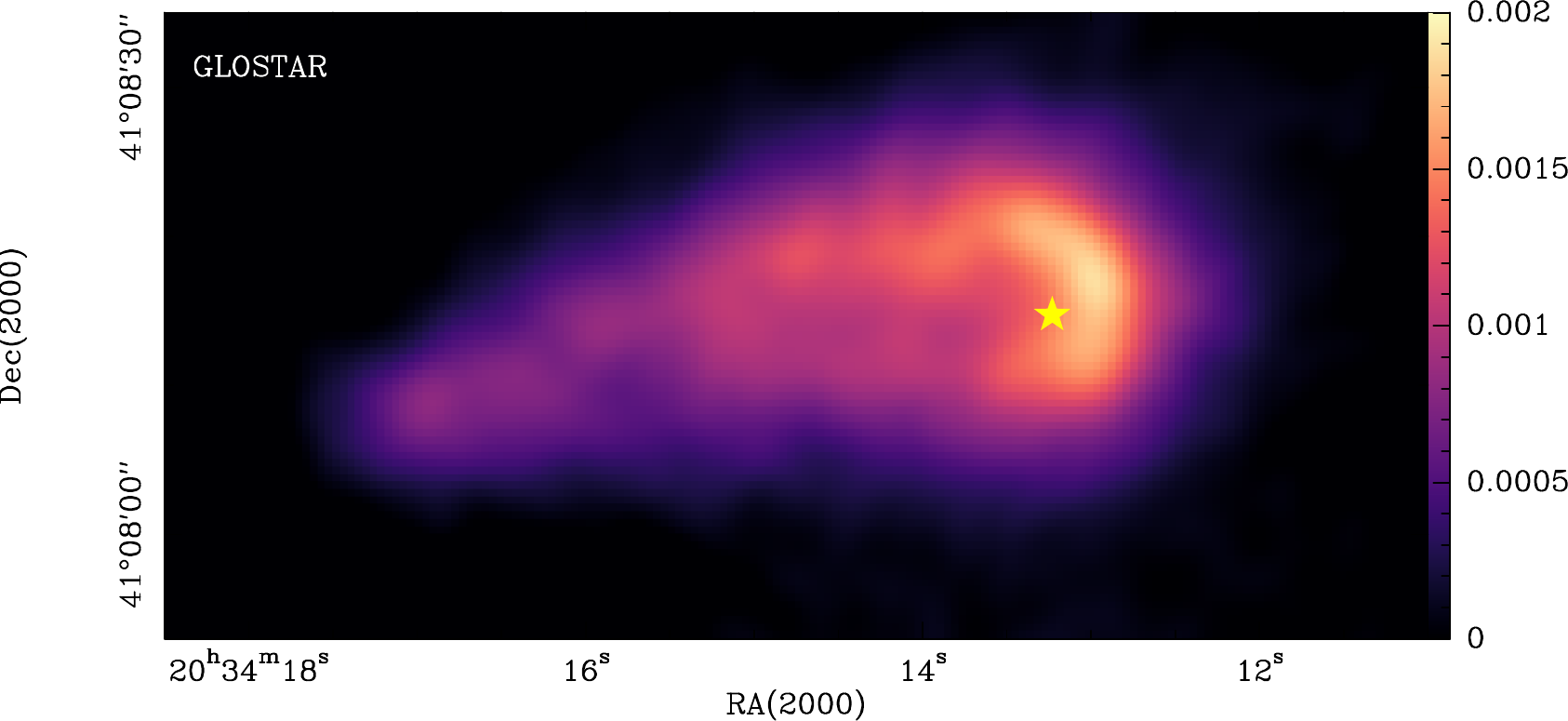}
\includegraphics [width=9cm, angle={0}]{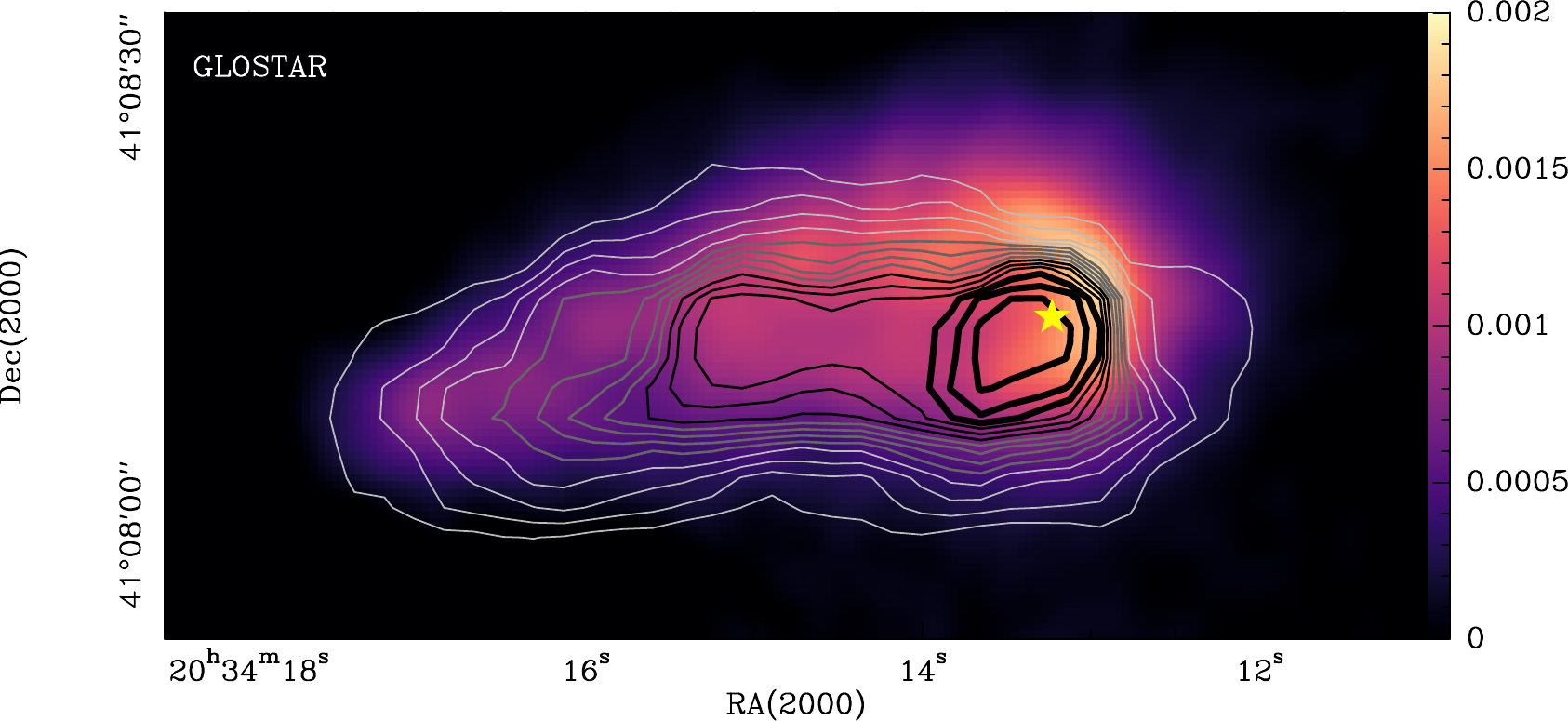}
\caption{ GLOSTAR 5.9 GHz images of proplyd \#7. The top image shows
  the GLOSTAR data only. The image scale is in Jy/beam at an angular
  resolution of 4$''$. The star indicates the position of the YSO. The
  lower panel displays the same radio image with contours of $^{13}$CO
  2$\to$1 emission overlayed. The contour lines have values of 5 to
  37.5 K km s$^{-1}$ in steps of 2.5 K km s$^{-1}$.}
\label{fig:glostar}
\end{center} 
\end{figure}

\begin{figure}[htbp]
\begin{center} 
\includegraphics [width=9cm, angle={0}]{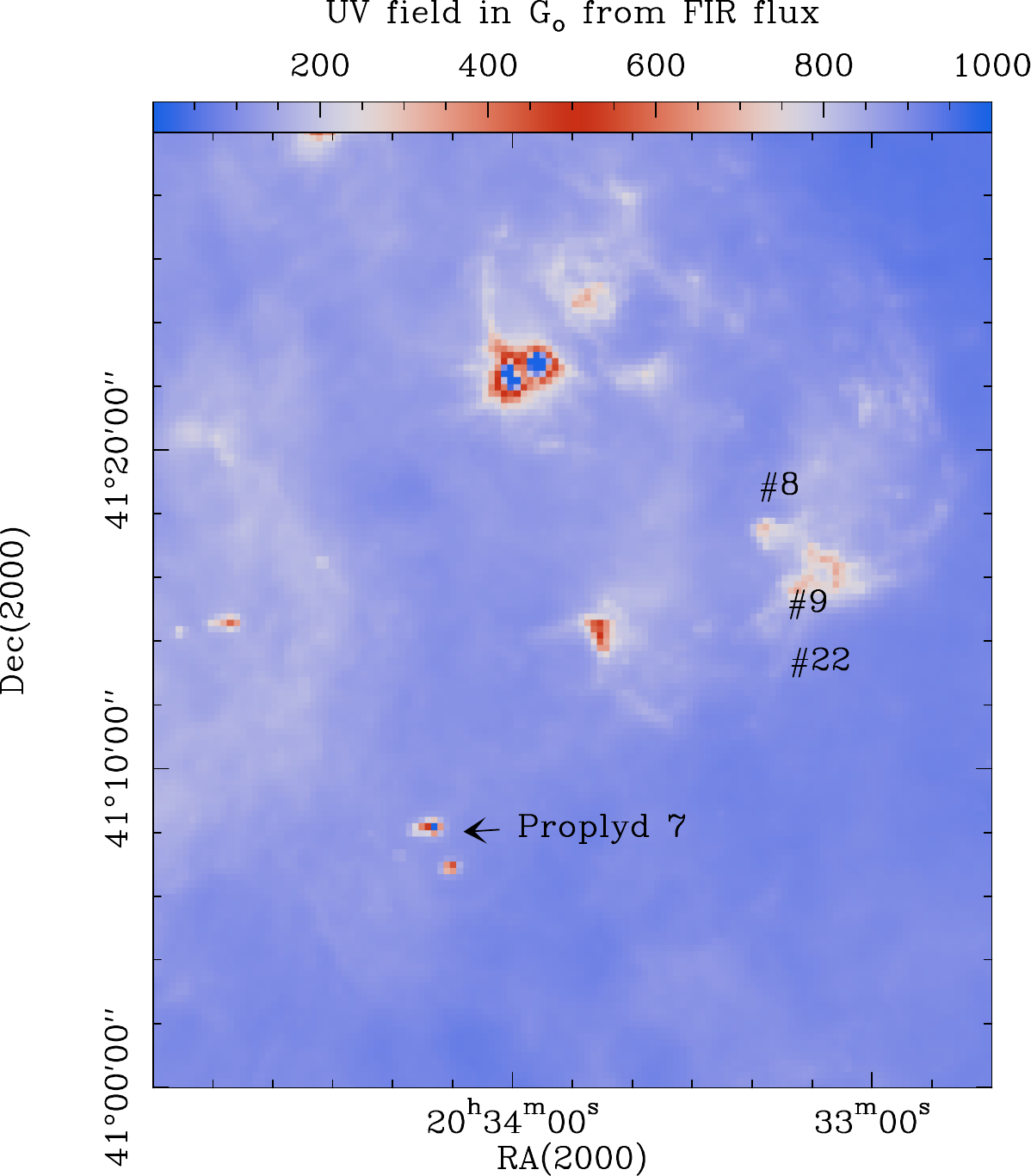}
\caption{FUV field around proplyd \#7. The FUV field in Habing units
  (G$_\circ$) is determined from the {\textit{Herschel}} 70 and 160
  $\mu$m fluxes on an angular resolution and on a grid of 11$''$.
  Proplyd \#7 is indicated by an arrow.}
\label{fig:UV}
\end{center} 
\end{figure}

\begin{table*} 
\caption{Values for the PDR modelling. }
\label{tab2}
\centering  
\begin{tabular}{c|c|ll|ll}     
\hline\hline
                          &            &                & &                & \\
Wavelength                & Instrument & Position 1     & & Position 2     & \\
                          &            & [K kms$^{-1}$] & [erg s$^{-1}$ cm$^{-2}$ sr$^{-1}$]  & [K kms$^{-1}$] & [erg s$^{-1}$ cm$^{-2}$ sr$^{-1}$] \\
\hline 
                           &          &                 & &                &\\     
63 $\mu$m     (\OI)        &  upGREAT &  5.3$\pm$0.5    & 5.8$\pm$0.6$\times$10$^{-4}$   & 1.5$\pm$0.5  & 1.6$\pm$0.6$\times$10$^{-4}$ \\
158 $\mu$m      (\CII)     &  upGREAT & 22.3$\pm$0.4    & 1.56$\pm$0.03$\times$10$^{-4}$ & 6.5$\pm$0.4  & 4.6$\pm$0.3$\times$10$^{-5}$ \\
1.3mm ($^{12}$CO 2$\to$1)  &  EMIR    & 101.7$\pm$0.6   & 1.27$\pm$0.07$\times$10$^{-6}$ & 30.0$\pm$0.6 & 3.74$\pm$0.07$\times$10$^{-7}$\\
\hline
                           &          & [MJy/sr]        &  & [MJy/sr] &  \\ 
70 $\mu$m                  &  PACS    &  12797          &  &  1148    &\\     
160 $\mu$m                 &  PACS    &  5402           &  &  1422    &\\     
\end{tabular} 
\tablefoot{All values were determined from data at an angular
  resolution of 14$''$. The two positions are indicated in
  Fig.~\ref{irac} and represent the location close to the YSO
  (position 1) and in the tail (position 2).}
\end{table*}

\begin{figure*}
\centering
\includegraphics[width=7.5cm]{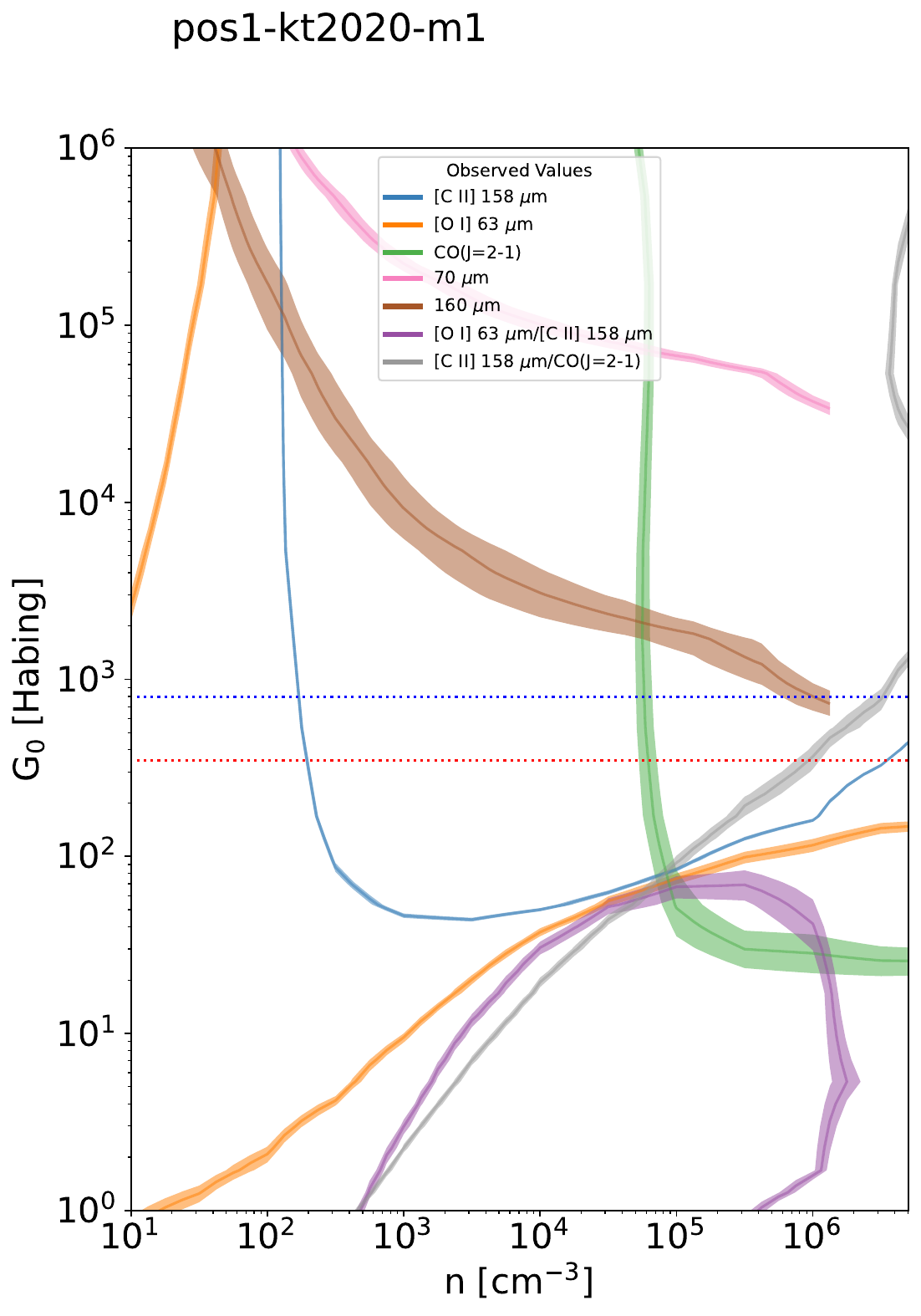}
\includegraphics[width=7.5cm]{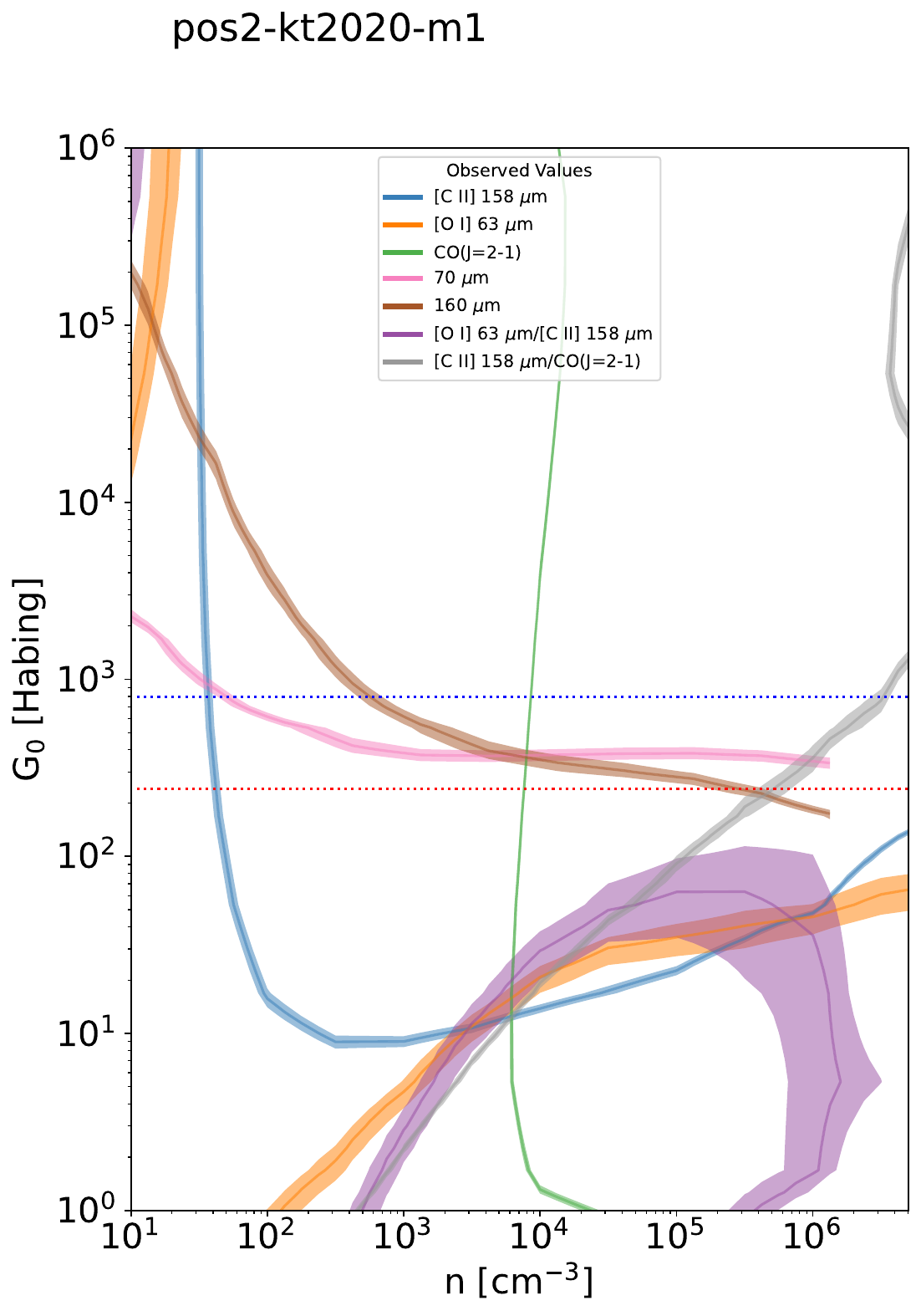}
\caption{PDR KOSMA-$\tau$ model results. The left (right) panel shows
  the diagnostic diagrams for UV-field and density, derived from the
  observed lines, line ratios, and FIR fluxes for positions 1 and 2 in
  proplyd \#7. The red and blue dashed lines indicate the UV field
  determined from the FIR {\textit{Herschel}} fluxes and from the
  stellar census, respectively, for positions 1 and 2. }
\label{fig:pdr}
\end{figure*}

\subsection{GLOSTAR observations} \label{subsec:glostar} 
Figure \ref{fig:glostar} displays the GLOSTAR image of proplyd~\#7 at
5.9 GHz, determined from the B+D VLA configuration at 4$''$ angular
resolution. The radio emission distribution is similar to what was
observed at 8.5 GHz and 22 GHz \citep{Sahai2012}. Firstly, the
emission forms a bright half-circle at the western head of proplyd
\#7. This appearance can have various explanations. It can be caused
by higher density due to external compression (radiation and wind from
the close-by OB-stars). However, it is also possible that the YSO is a
massive star and created a compact \HII\ region, where the expanding
ionised gas is compressed from the inside and forms a denser and,
thus, brighter shell in the west while towards the east, dense
molecular gas forms a barrier. This scenario would well explain the
half-circle morphology. No very compact emission was detected in the
GLOSTAR B-configuration which is sensitive to structures up to 4$''$,
that correspond to 0.03 pc at a distance of 1.4 kpc. This could imply
that we do not observe an ultra-compact \HII\ region (sizes of $<$0.1
pc), but moreover a more evolved, compact \HII\ region that is
typically 0.1 to 0.5 pc in size, depending on density.

The large-scale 5.9 GHz emission can have several reasons. It is clear
that enough EUV photons are required to generate the ionised hydrogen
for the free-free emission.  Proplyd \#7 could be embedded in an
environment with a sufficient large fraction of ionised gas. We
checked the Cygnus mapping at 1$'$ resolution of the Canadian Galactic
Plane survey \citep{Taylor2003} at 1420 MHz and indeed reveal that
there is an enhanced level of $\sim$12~mJy emission around proplyd
\#7. Figure 8 in \citet{Setia2003} shows also a higher emission level
around proplyd \#7 (their source 118). Another argument for external
illumination is that all other proplyd-like objects in this area also
shine at 5.9 GHz, even those who clearly do not have an internal YSO.

Nevertheless, it is also possible that EUV photons stem from inside.
In this case, the central YSO must be a massive star (see above) with
an internal \HII\ region and the material east of the bright emission
spot must be very clumpy so that the EUV photons can channel through
the molecular clump. Proplyd \#7 has a fragmented structure, this
becomes obvious in the overlay with line integrated $^{13}$CO emission
(lower panel of Fig. \ref{fig:glostar}) and in the channel maps
displayed in Fig. \ref{fig:IRAM2}. However, higher angular resolution
molecular line observations are necessary to better reveal the
structure.

From the drop in flux between 8.5 GHz and 22 GHz, \cite{Sahai2012}
argued in favour of non-thermal emission (Fig. \ref{fig:pr7sed}
displays the radio SED). However, the observed flux density at 5.9 GHz
is at around 49~mJy and, thus, it is significantly lower than what
would be expected from purely non-thermal emission, indicating an
optically thick regime consistent with a thermal \HII\ region;
meanwhile, the 22 GHz emission is in the optically thin part of the
SED. We note that the flux density at 22 GHz is a lower limit because
these observations were performed in the C-configuration and even
though an UV-taper was applied to arrive to a similar beam as for the
8.5 GHz observations, the shorter baselines are missing at 22 GHz so
that extended emission gets lost.  In summary, from the overall
emission distribution (Fig. \ref{fig:pr7sed}) and the comparison
between the absolute values of radio emission, we deduced that we
observe typical free-free emission from a thermal compact
\HII\ region. This conclusion is similar to what was stated in
\cite{Isequilla2019}, described with a turnover frequency of a few
GHz.

\section{Analysis} \label{sec:analysis} 
\subsection{The external FUV field of proplyd \#7} \label{subsec:fuv} 
A central question in revealing the nature of proplyd~\#7 is to assess
how much of the observed emission in all line tracers is caused by
internal and external radiation. Given that proplyd \#7 is oriented
towards the centre of the Cyg OB2 association, its morphology was most
likely influenced by external stellar feedback. \citet{Pereira2007}
and \citet{Wright2012} showed in their H$_\alpha$ maps that the
northern edge of proplyd \#7 is brighter than the southern part,
suggesting an influence from the central O-stars of Cyg OB2 (mainly
\#8, 9, and 22 from the notation of \citet{Schulte1958}. However, the
overall orientation of proplyd \#7 is west-east and points mostly
towards the star \#720 (notation from \citealt{Massey1991}; see our
Fig. \ref{overview}).  Figure~\ref{fig:UV} \citep{Schneider2016a}
displays the FUV field, obtained by translating the observed
{\textit{Herschel}} 70 and 160 $\mu$m fluxes into a FUV field,
assuming all incoming radiation is re-radiated in the FIR. We note
that it is clear that a FUV field can only be determined in this way
when there is enough material so that all FUV photons can be absorbed
in a PDR. Otherwise it provides a lower limit only. We obtained values
of 352 and 239 G$_\circ$ for positions 1 and 2, respectively and a
high field of 1737 G$_\circ$ at the location of the YSO. From the
stellar census\footnote{Note: the derived FUV field is an upper limit
because it is assumed that all stars are in the plane of the sky and
that there is no extinction of the UV field by dust and gas of the
ISM.}, \citet{Schneider2016a} derived an upper limit of $\sim$1000
G$_\circ$ averaged over proplyd \#7. Because at the position of the
YSO we find a significantly higher value, it is obvious that there is
additional heating (potentially through an internal PDR).

\subsection{PDR (photodissociation region) modelling} \label{subsec:pdr} 
We employed KOSMA-$\tau$ \citep{Roellig2006,Roellig2022}, a
well-tested \citep{rollig2007photon} PDR model to derive local
physical conditions from the observed \OI, \CII\, and CO 2$\to$1 line
intensities and ratios, and FIR 70 and 160 $\mu$m
fluxes. \\ KOSMA-$\tau$ numerically computes the energetic and
chemical balance in a spherical cloud with a density gradient similar
to a Bonnor-Ebert sphere that is externally irradiated. The whole
region is represented by an ensemble of such spherical clumps. While
the model allows for superpositions of different clumps with a size
spectrum, we only used the simplified approach with identical clumps
sizes.  The full numerical computation scheme of KOSMA-$\tau$ involves
three steps.

First, the continuum radiative transfer code MCDRT
\citep{szczerba1997iras} was used to compute the thermal balance of
all dust components as well as the FUV radiative transfer within the
model cloud, and the emergent continuum radiation. We employed the
dust model 7 from \citet{weingartner2001dust}, which assumes that the
impact of the gas on the dust temperature is negligible. The MCDRT
output was then used as input for the second step, where KOSMA-$\tau$
computes the chemical structure and temperature of the gas in
equilibrium. The result are radial chemical and temperature profiles
of the model clump. In a third step, they are used to perform the line
radiative transfer computations, giving the spectral emission of the
model cloud for comparison with observations \citep{gierens1992uv}. In
the model runs, we applied the most recent CO self shielding functions
by \citet{visser2009photodissociation} and assume a Doppler line width
according to the line width-and-size relation by
\citet{larson1981turbulence}. The photoelectric heating is estimated
according to \citet{weingartner2001photoelectric}. The formation of
H$_2$ on grain surfaces follows
\citet{cazaux2002molecular,cazaux2004h2,cazaux2010erratum} and
formation on PAH surfaces is suppressed. We applied the UMIST Database
for Astrochemistry chemical network \citep{mcelroy2013umist},
including all isotopic reaction variants including $^{13}$C and
$^{18}$O isotopes. More details on the model and how it is applied to
SOFIA data can be found in \citet{Schneider2018,Schneider2021}.

In Table~\ref{tab2}, we give the values of the \OI, \CII, and CO
2$\to$1 line intensity for two positions where we performed the
modelling processes indicated in Fig.~\ref{irac}. Positions 1
($-$4$''$,0$''$) characterises the peak of \OI\ emission and is close
to the location of the YSO, while position 2 (36$''$,0$''$) is in a
more quiescent region in the tail of proplyd \#7. The upper limit for
the mass for both positions in one beam is around 2 M$_\odot$,
determined from the {\textit{Herschel}} column density map. Thus, we
used the KOSMA-$\tau$ models that were calculated for a clump mass of
1 M$_\odot$; however, we note that the model results depend only
weakly on the clump mass.

The \OI\ line integrated (over the whole velocity range) intensity is
5.8$\times$10$^{-4}$ erg s$^{-1}$ sr$^{-1}$ cm$^{-2}$ and
1.6$\times$10$^{-4}$ erg s$^{-1}$ sr$^{-1}$ cm$^{-2}$ for positions 1
and 2, respectively. These values are lower than what was observed for
the proplyds in Carina (4.5$\times$10$^{-3}$ erg s$^{-1}$ sr$^{-1}$
cm$^{-2}$) and Orion (0.36 and 2.1 erg s$^{-1}$ sr$^{-1}$ cm$^{-2}$)
by \citet{Champion2017}.

Figure~\ref{fig:pdr} presents the parameter space of gas density and
FUV radiation field derived from the KOSMA-$\tau$ model (2020),
implemented in the PDR Toolbox \citep{Pound2023} as `non-clumpy'. The
intersection of the observed line intensities and diagnostic line
ratios delineates the most probable values for both the gas density
and the FUV field strength.  From the plots, it becomes obvious that
the PDR model does not deliver one simple solution for the density and
UV-field for the two positions. For position 1, the FIR line emission
and ratios indicate a FUV field $<$100 G$_\circ$) that is lower than
the one estimated from the FIR flux-based method. The
{\textit{Herschel}} flux values of 70 and 160 $\mu$m cover a parameter
space with much higher FUV field. The derived gas density has values
of the order of 10$^5$ cm$^{-3}$, which is consistent with
expectations for dense PDRs.

For position 2, the FIR lines point towards an even lower FUV field
(around 20 G$_\circ$) and a density of 10$^4$ cm$^{-3}$, while the
{\textit{Herschel}} flux values and the $^{12}$CO 2$\to$1 line cross
approximately the estimated field of around 240 G$_\circ$ at a density
of 10$^4$ cm$^{-3}$. In addition, the PDR model delivers a surface
temperature of around 150 to 200 K.

The inconsistency between the fit results for the line and continuum
data suggests that the chemical and thermal structure of proplyd \#7
are governed by several processes and not by FUV radiation
alone. Considering only the effect of photo-dissociation, the actual
PDR chemistry is traced through the \OI\ and \CII\ lines where the
observations suggest a relatively low FUV field. In contrast the
thermal dust emission traces the total energy input into the ISM,
including compression, shocks, magneto-hydrodynamic (MHD) dissipation,
and radiation outside of the FUV wavelength range. The fact that the
continuum levels suggest much higher radiation fields in the fit
indicates that those non-FUV processes are dominant in proplyd \#7 (at
least at the position of the YSO), so that it cannot be described as a
simple PDR. Nevertheless, the model is useful to constrain the gas
density and assess the pure FUV effect.

\begin{table} 
\caption{
Physical overall averaged properties of proplyd \#7. 
    }
\label{tab-phys}
\centering  
\begin{tabular}{c|c|ll|ll|c}     
\hline\hline
                 &                &                &                 &       \\
                 & N$^a$          & N(H)$^b$    & N(H${_2}$)$^c$ & Mass  \\
                 & [cm$^{-2}$]    & [cm$^{-2}$]    & [cm$^{-2}$]     & [M$_\odot$] \\
\hline 
                 &                       &                &                 &  \\     
{\textit{Herschel}}   &                       &                &  9.0$\times$10$^{21}$  &  19.1 \\
CO 2$\to$1       &  5.4$\times$10$^{15}$ &                &  6.1$\times$10$^{21}$  &  16.3 \\
\CII\ 158 $\mu$m &  3.7$\times$10$^{17}$ & 3.1$\times$10$^{21}$  &                 &   3.9 \\
\OI\ 63 $\mu$m   &  1.0$\times$10$^{18}$ & 5.5$\times$10$^{21}$  &                 &   1.8 \\
\end{tabular} 
\tablefoot{Most likely values for the column densities and masses of proplyd \#7, derived in Appendix~\ref{appendix-cal}.
\tablefoottext{a}{Column density of the individual species (CO, \CII, and \OI).}
\tablefoottext{b}{Atomic hydrogen column density.}
\tablefoottext{c}{Molecular hydrogen column density.}
}
\end{table}

\subsection{The physical properties and star-formation history of proplyd \#7} \label{subsec:props}  
We used the CO data to determine the excitation temperature and column
densities (N$_{\rm CO}$ and N(H$_2$)) across proplyd \#7 as well as to
estimate the total molecular gas mass. The FIR \OI\ 63 $\mu$m and
\CII\ 158 $\mu$m lines were used to derive the mass of the atomic gas
component associated with the PDR of proplyd \#7.  Detailed
calculations are given in Appendix~\ref{appendix-cal} and we also
summarise these results in Table~\ref{tab-phys}. We note that all
values are an approximation and have significant uncertainties (see
Appendix~\ref{appendix-cal}). The total mass is an important property
because it determines its evolution and lifetime and, thus, the total
time available for accretion by the star(s) that might form inside it
and for any further stars that might subsequently form.

The molecular mass derived from CO observations is 16.3 M$_\odot$,
which is comparable to the 19.1 M$_\odot$ obtained from the
{\textit{Herschel}} column density map that includes both atomic and
molecular hydrogen (see column 4 in Table~\ref{tab-phys}). In
contrast, the atomic gas mass is significantly lower, with 1.8
M$_\odot$ and 3.9 M$_\odot$, derived from \OI\ and \CII\ emission,
respectively. Thus, the atomic PDR associated with proplyd~\#7
contributes only a minor fraction to the total mass. The mass-loss
rate is estimated using Eq.~3 from \citet{Schneider2016a}, adopting
the same values for the external UV photon flux, the equivalent
radius, and the distance to the central stars of Cygnus OB2. This
yields a mass-loss rate of $9.65 \times 10^{-5}$~M$_\odot$ yr$^{-1}$,
at least a factor 100-1000 larger than the typical values for the
Orion proplyds \citep{Henney1999,Sheehan2016,Winter2019,
  Ballering2023,Aru2024,Boyden2025}. The photo-evaporation time
(t$_{photo}$), defined as the time required to completely
photo-evaporate the object (assuming constant density but a variable
radius), is calculated using Eq.~4 in \citet{Schneider2016a},
resulting in t$_{photo}$ = 1.62$\times$10$^5$ yrs. We adopted the UV
exposure time for proplyd~\#7 as t$_{expos}$ = 4.52$\times$10$^6$ yrs,
based on Table~A.2 of \citet{Schneider2016a}. The free-fall time for
an isothermal, gravitational collapse is computed using the standard
expression t$_{ff} = \sqrt{(3 \pi /(32 \, G \, n \,m_H\, \mu)}$, where
$G$ is the gravitational constant, $n$ = 4.4$\times$10$^3$ cm$^{-3}$
is the average particle number density, $m_{H}$ is the mass of a
hydrogen atom, and $\mu$ = 2.33 is the mean molecular weight
accounting for the presence of heavier elements. We obtain a value of
t$_{ff}$ = 5.2$\times$10$^5$ yrs.  The photo-evaporation timescale is
relatively short, 1.62$\times$10$^5$ yrs, consistent with values
derived for the proplyds in Orion \citep{Johnstone1998}. In contrast,
the free-fall timescale is more than twice as long, indicating that
proplyd \#7 is likely to be dispersed by external irradiation before
it can undergo gravitational collapse and form more stars. The
lifetime of proplyd \#7 is probably even shorter if we consider the
feedback effects (radiation and wind) from the internal YSO.

\section{Conclusions: Proplyd \#7 as an embedded protoplanetary disc or an irradiated globule} \label{sec:discuss} 

Our millimetre (mm), FIR, and radio observations of proplyd \#7 are
limited by angular resolution, ranging from 4$''$ to 14$''$, which
hampers the ability to disentangle emission from a potential
protoplanetary disc and the molecular envelope. The embedded YSO
within proplyd \#7 may still be accreting mass from the surrounding
gas reservoir. This scenario contrasts with that of typical Orion
proplyds, where the disc is actively losing mass due to external
photo-evaporation.  Despite these limitations, we detected outflow
signatures in the CO, \OI, and \CII\ lines near the location of the
YSO. These emissions could originate from two plausible scenarios:
\begin{itemize} 
\item(i) a protoplanetary disc associated with the central star, either a massive star or a T Tauri star. \\ 
\item(ii) gas ablated from the cavity walls by the radiation and stellar winds of a massive central star in the absence of a disc.
\end{itemize} 
Both scenarios are possible; however, scenario (ii) is more probable,
mostly because of the radio continuum observations that point toward
the existence of a thermal \HII\ region and stellar classifications in
the literature \citep{Comeron2002,Isequilla2019}, which suggest that
the central object is an OB star. In this case, the \CII\ and
\OI\ lines may be shock-excited, explaining why the observed line
intensities and FIR continuum fluxes deviate from the KOSMA-$\tau$ PDR
model predictions. The CO emission is likely entrained in the ionised
outflow. This would explain the contradiction why the FIR continuum
emission in proplyd \#7 is reasonably well reproduced by an external
UV radiation field of a few hundred G$_0$ (at least for position 2),
but the observed \OI\ and \CII\ line intensities are better matched by
models assuming a lower FUV field.  However, the situation is more
complex. We observe the same morphology in dust, radio continuum (that
we interpret as thermal), and CO, which is inconsistent because on one
hand, EUV photons need to penetrate for the free-free emission;
however, FUV shielding is required for the formation of CO. Having
both in the same structure is only possible if we have a very clumpy
medium with dense, self-shielding clumps and a highly fragmented
structure with many channels for UV penetration in between. However,
in this case, we tend to run into problems for the calculation of the
FUV field. The FUV estimate from the FIR emission assumes that all FUV
photons are absorbed; thus, accordingly, we cannot have too many
leaking photons (i.e. there is a margin of about a factor 2 when
looking at the numbers). However, the EUV extinction should exceed the
FUV extinction. Therefore, it is unclear how we can manage to still
have enough EUV photons to generate all the ionised hydrogen required
for free-free emission. This is only possible if proplyd \#7 was
indeed found to be embedded in a significantly ionised medium, which
is not fully the case (Sect. \ref{subsec:glostar}).

Overall, the external irradiation from nearby OB stars in the Cyg OB2
cluster may have had only a moderate impact on proplyd \#7. The object
has been exposed to UV radiation for approximately 4.5 10$^6$ yrs and
has persisted over this timescale.  It remains uncertain whether the
formation of the central star(s) resulted from radiative and
wind-driven gas compression by the OB stars or from spontaneous
gravitational collapse of an isolated molecular clump. The Jeans
mass\footnote{This is the critical mass M$_J$ above which a gas cloud
becomes gravitationally unstable and begins to collapse under its own
gravity and is calculated by M$_J = \left( \frac{5 \, k_b \, T}{G \,
  \mu \, m_H}\right)^{3/2} \, \left(\frac{3}{4\,\pi \,n}\right)^{1/2}
\approx 1 \,M_\odot \,\left(\frac{T}{10 K}\right)^{3/2} \,
\left(\frac{n}{10^4 cm^{-3}}\right)^{-1/2}$.}  is a few M$_\odot$ for
typical temperatures of 10-20 K and densities 10$^3$ to 10$^4$
cm$^{-3}$ and, thus, much lower than the current gas mass of $\sim$20
M$_\odot$, as derived from CO and dust (the initial cloud mass was
likely higher). In contrast, the atomic gas mass (estimated from
\OI\ and \CII\ emission) is relatively low, on the order of a few
M$_\odot$.  From a simulation viewpoint, the ongoing star formation
within proplyd \#7 has not necessarily been triggered externally
(e.g. by a shock front preceding a D-type ionization front). Smooth
particle hydrodynamic simulations presented in \citet{Bisbas2011},
which model the propagation of ionizing radiation and the resulting
dynamical evolution of a molecular clump, show in their Fig.~12 a
parameter space of ionizing flux and clump mass that distinguishes
between regimes of triggered star formation and clump
dispersal. Proplyd \#7 has an upper limit for the ionizing flux of
$4.8\times10^{10}$~cm$^{-2}$, estimated from a geometric dilution
factor of $1/(4\pi d^{2})$ at a distance, $d$, of 5.9~pc from the
central Cyg~OB2 cluster, assuming a photon emission rate of
$10^{50.3}$~s$^{-1}$ \citep{Schneider2016a}. With a mass of
$\sim20~M_\odot$ (Table~\ref{tab-phys}), this places proplyd \#7
within the regime where the clump is expected to be dispersed rather
than to undergo externally triggered star formation through mechanisms
such as radiatively driven implosion \citep{Lefloch1994}.  This is
supported by arguments of timescales. Proplyd \#7 is unlikely to
survive much longer because given its photo-evaporation timescale of
1.62$\times$10$^5$ yrs and a free-fall time of 5.2$\times$10$^5$ yrs,
it is improbable that further star formation will occur within this
object.

Although significant progress has been made in understanding the
nature of proplyd~\#7, further observations are required to address
the key question of whether we are observing a disc around a massive
star within a globule (a remarkable discovery if confirmed) that is
driving a protostellar CO outflow. NIR spectroscopy and imaging of
H$_2$ and Br$\gamma$ emission from the central object would aid in
determining the spectral type of the potential massive star. However,
this is most likely not possible because the spectrum of the central
star is strongly veiled, as demonstrated in \citet{Guarcello2014}. One
could use for example the \textit{James Webb }Space Telescope (JWST)
to observe CO vibrational bands to trace the hot and cold gas and
light hydrocarbons. H$_2$ NIR (NIRSpec) observations could provide a
much better understanding of the distribution of the molecular gas.
New FIR cooling line observations are no longer possible following the
shutdown of SOFIA; however, the \OI\ 63~$\mu$m line emission, together
with \CII\ emission and hydrogen column densities, can be incorporated
into shock models such as the Paris-Durham model
\citep{Godard2019}. However, this analysis lies beyond the scope of
the present work. Such modelling would reveal whether the relatively
strong \OI\ and \CII\ emissions originate from protoplanetary disc
environments where shocks and high-energy processes occur. These
processes could include accretion shocks arising from gas infall onto
the central star, heating of the inner disc regions that induces
shocks, or jet-driven shocks generated either by outflows from the
central star or by interactions within the disc that propagate into
the surrounding medium. On the other hand, \OI\ emission arising from
an atomic jet linked to the disc-envelope-star system is predicted to
emerge from high-velocity ($\ll$100 km s$^{-1}$) dissociative J-type
shocks that hit the dense surrounding gas \citep{Hollenbach1985,
  Hollenbach1989}.  In contrast, the observed redshifted \OI\ emission
exhibits relatively low projected velocities of $\sim$2–5 kms$^{-1}$.

In summary, proplyd~\#7 retains unresolved characteristics and
apparent contradictions that warrant further investigation. It remains
uncertain whether proplyd~\#7 truly merits its designation as a
protoplanetary disc or whether it is instead a compact, small globule
undergoing internal star formation.  However, the study presented here
sets constraints on the timeline of the evolution of this object and
highlights observational discrepancies that future higher angular
resolution studies have to solve.

\begin{acknowledgements}
We thank the anonymous referee for very useful comments.  This study
was based on observations made with the NASA/DLR Stratospheric
Observatory for Infrared Astronomy (SOFIA). SOFIA is jointly operated
by the Universities Space Research Association Inc. (USRA), under NASA
contract NNA17BF53C, and the Deutsches SOFIA Institut (DSI), under DLR
contract 50 OK 0901 to the University of Stuttgart. (up)GREAT is a
development by the MPI f\"ur Radioastronomie and the
KOSMA/Universit\"at zu K\"oln, in cooperation with the DLR Institut
f\"ur Optische Sensorsysteme.  This work partially uses data from the
GLOSTAR survey, whose database is available at
\href{http://glostar.mpifr-bonn.mpg.de}{http://glostar.mpifr-bonn.mpg.de}
and is supported by the MPIfR, Bonn.  N.S. acknowledges support by the
BMWI via DLR, Projekt Number 50OR2217.  S.K. acknowledges support by
the BMWI via DLR, project number 50OR2311.  This work was supported by
the CRC1601 (SFB 1601 sub-project A6, B2) funded by the DFG (German
Research Foundation) – 500700252.  S.D. acknowledges support by the
International Max Planck Research School (IMPRS) for Astronomy and
Astrophysics at the Universities of Bonn and Cologne.
S.A.D. acknowledges the M2FINDERS project from the European Research
Council (ERC) under the European Union's Horizon 2020 research and
innovation programme (grant No 101018682). \\ G.N.O.L. acknowledges
the financial support provided by Secretaría de Ciencia, Humanidades,
Tecnología e Innovación (Secihti) through grant CBF-2025-I-201.\\ We
thank the director of IRAM, Karl Schuster, for providing us with
Director’s Discretionary Time to observe proplyd \#7 with the IRAM 30m
telescope.
\end{acknowledgements}

\bibliographystyle{aa} 
\bibliography{bibliography} 

\begin{appendix}

\section{Complementary plots} \label{appendix-plots}
\begin{figure*}[htbp]
\begin{center} 
\includegraphics [width=8cm, angle={0}]{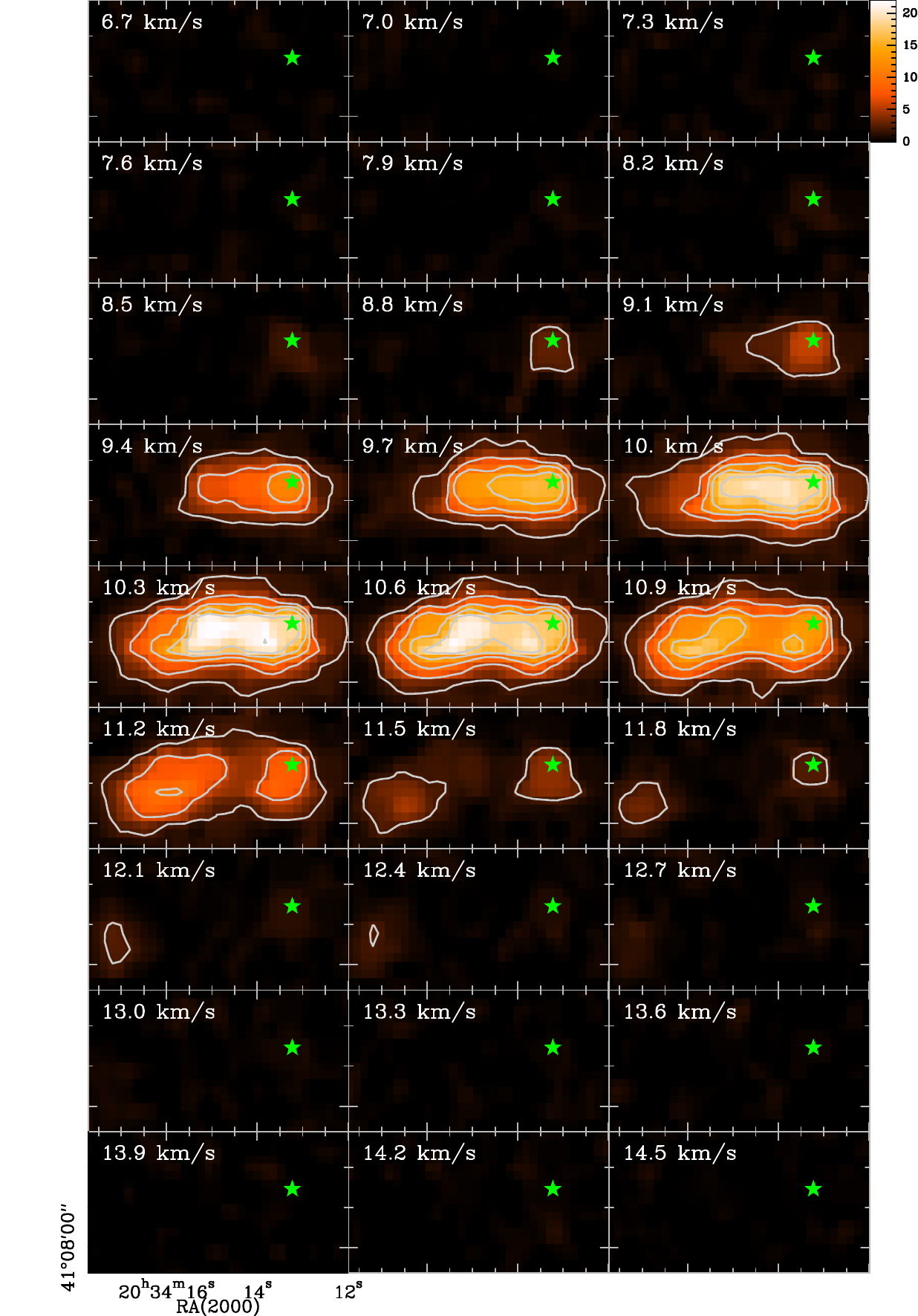} 
\includegraphics [width=8cm, angle={0}]{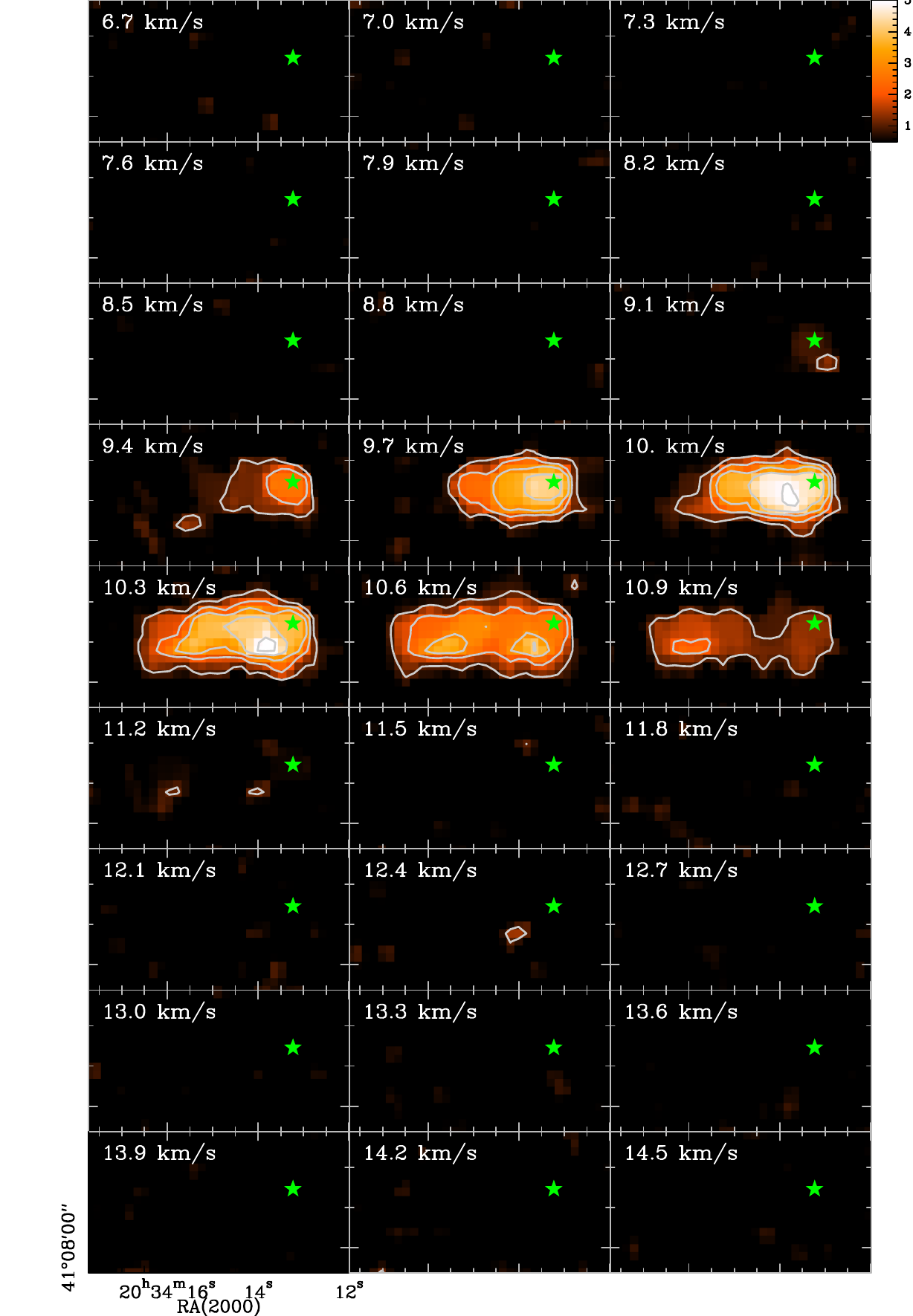} 
\caption{Channel maps of CO emission. The left (right) panel shows a
  channel map of $^{13}$CO (C$^{18}$O) 2$\to$1 emission from 0 to 22
  (0 to 5) K km s$^{-1}$ in steps of 4 km s$^{-1}$ (0.3 km
  s$^{-1}$). The green star indicates the position of the protostar.}
\label{fig:IRAM2}
\end{center} 
\end{figure*}

Figure \ref{fig:IRAM2} displays channel maps of the $^{13}$CO and
C$^{18}$O 2$\to$1 emission of proplyd \#7. The optically thin
C$^{18}$O has peak emission around 10 km s$^{-1}$ close to the
position of the YSO, while the marginally optically thin $^{13}$CO
also extends further east. At higher velocities, above 11 km s$^{-1}$,
proplyd \#7 fragments into two separated clumps.

\begin{figure}[htbp]
\begin{center} 
\includegraphics [width=8cm, angle={0}]{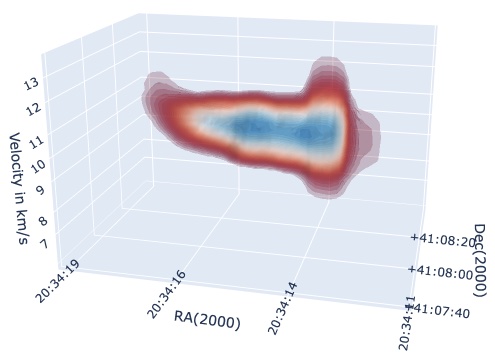} 
\caption{
3D Position-velocity cut of proplyd \#7 in $^{12}$CO 2$\to$1. An interactive version is available online. 
}
\label{fig:3D-PV}
\end{center} 
\end{figure}

Figure \ref{fig:3D-PV} displays one viewpoint of a 3D
position-velocity cut of $^{12}$CO 2$\to$1 emission. The outflow with
red- and blue-shifted emission is clearly visible at the position of
the YSO (RA(2000) = 20$^h$34$^m$14$^s$, Dec(2000) =
41$^\circ$8$'$). In addition, proplyd~\#7 appears slightly distorted,
with the tip of its tail exhibiting higher red-shifted
velocities. This may indicate rotational motion, with the head and
tail rotating at different speeds, a behaviour previously observed in
a Cygnus globule \citep{Schneider2021} using the \CII\ 158 $\mu$m
line.

\begin{figure}[htbp]
\begin{center} 
\includegraphics [width=8cm, angle={0}]{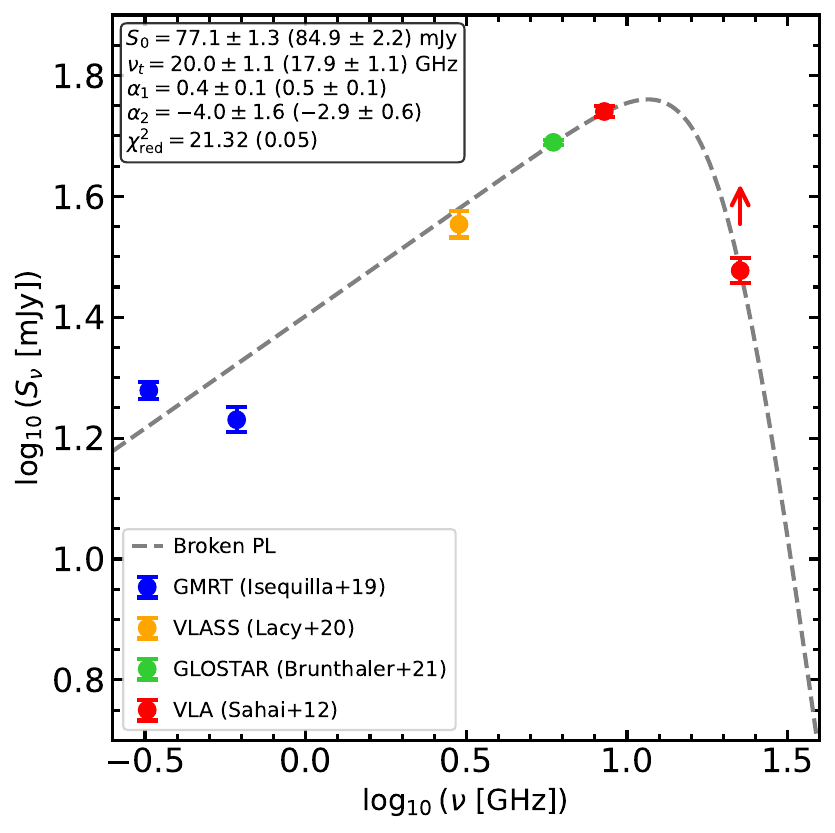} 
\caption{Radio SED of proplyd 7. The dashed line corresponds to the
  best fit of a broken power law (PL; see Eq. \ref{eq:broken_pl}),
  with the fit results given in the upper-left panel and in brackets
  the fit results for the same function, but excluding the first data
  point. The red arrow on the last data point indicates that this flux
  value is likely underestimated.}
\label{fig:pr7sed}
\end{center} 
\end{figure}

Figure~\ref{fig:pr7sed} presents the radio SED of proplyd~\#7,
incorporating values from the literature along with our GLOSTAR
measurement. The corresponding flux densities are listed in
Table~\ref{tab:fluxvalues}. The fitted broken power law follows the
form,
\begin{equation}
\label{eq:broken_pl}
S(\nu) = S_0 \left[\frac{\nu}{\nu_t}\right]^{\alpha_1} \left[1 +
  \left(\frac{\nu}{\nu_t}\right)^{\alpha_1 - \alpha_2}\right]^{-1}
.\end{equation} Here, $S_0$ corresponds to the peak flux density,
$\alpha_1$ and $\alpha_2$ are the low- and high-frequency spectral
indices respectively, and $\nu_t$ is the transition frequency.  The
overall shape of the SED is consistent with a compact thermal
\HII\ region. Only data with comparable angular resolution, for which
the beam can be assumed to be fully filled, have been
included. Although differences in resolution introduce some
uncertainty, the SED indicates peak emission of $\sim$80 mJy at 18-20
GHz, consistent with a partially optically thick regime at lower
frequencies, where $S_\nu \propto \nu^\alpha$ with $\alpha < 2$. At
22~GHz, the emission should lie in the optically thin regime,
following $S_\nu \propto \nu^{-0.1}$. We argue here that the low
values for $\alpha_2$, constrained only by a single data point at 22
GHz, are due to the use of the C-configuration which misses the short
baselines and thus extended emission.

\begin{table}[htbp]
\centering
\caption{Flux density values of Proplyd \#7.}
\label{tab:radio_obs}
\footnotesize 
\begin{tabular}{ccccl}
\hline
\hline
$\nu$ & $S_\nu$ & Resolution & Reference \\
(GHz) & (mJy) & (") & \\
\hline
0.325 & $19.0 \pm 0.6$ & $7.8$ & \citet{Isequilla2019} \\
0.610 & $17.0 \pm 0.8$ & $7.6$ & \citet{Isequilla2019} \\
3.0   & $35.8 \pm 1.8$  & $2.5$  & \citet{Lacy2020} \\
5.9   & $48.9 \pm 0.5$ & $4.0$   &  this work \\
8.5   & $55.0 \pm 1.2$ & $3.2$ & \citet{Sahai2012} \\
22.5  & $30.0 \pm 1.4$ & $3.2$ & \citet{Sahai2012} \\
\hline
\label{tab:fluxvalues}
\end{tabular}
\tablefoot{Flux densities and uncertainties from integrated measurements. Resolution values represent the synthesised beam FWHM.}
\end{table}

\section{Calculation of physical properties} \label{appendix-cal}
\begin{figure}[htbp]
\begin{center} 
\includegraphics [width=8cm, angle={0}]{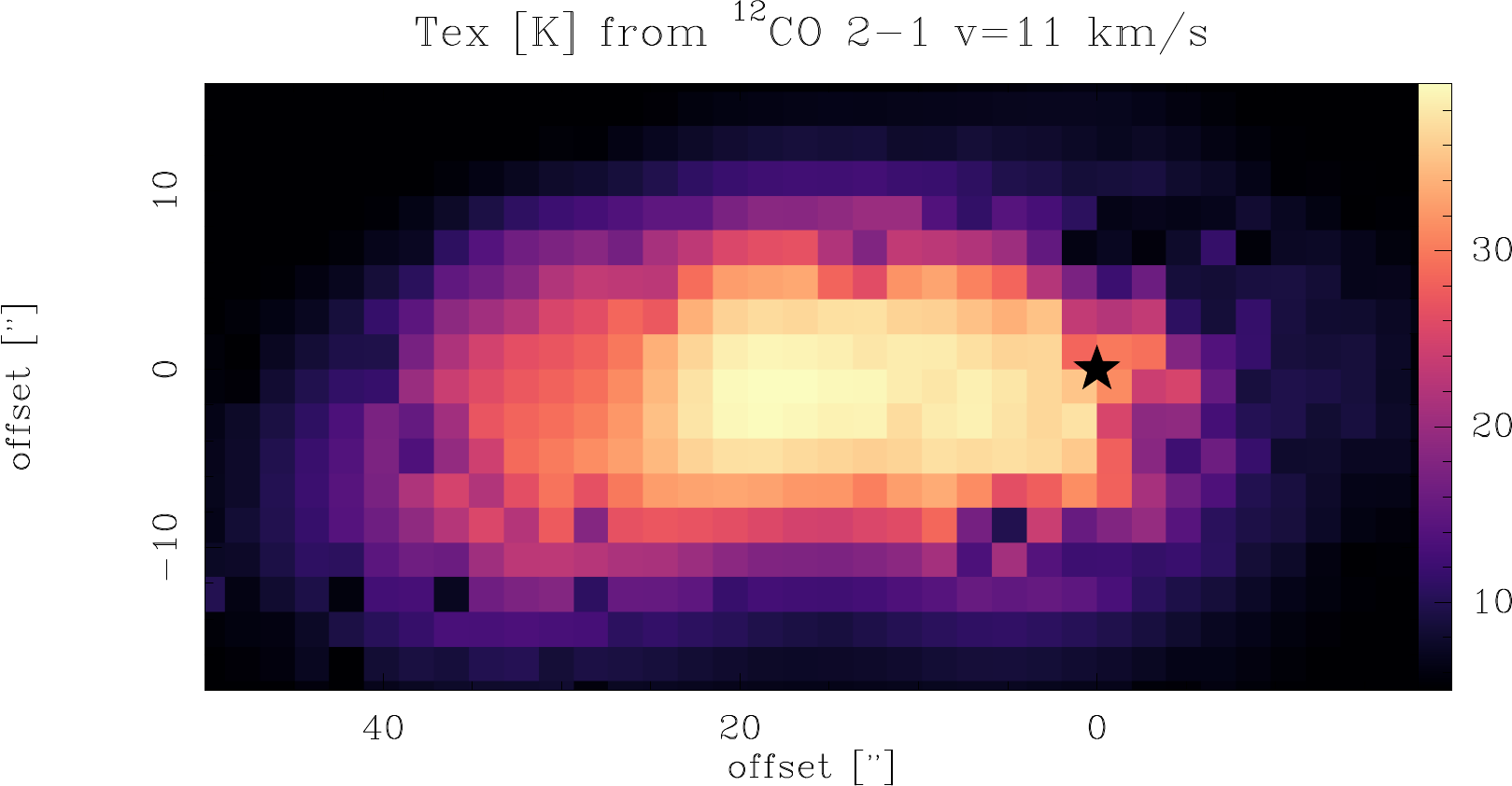} 
\includegraphics [width=8cm, angle={0}]{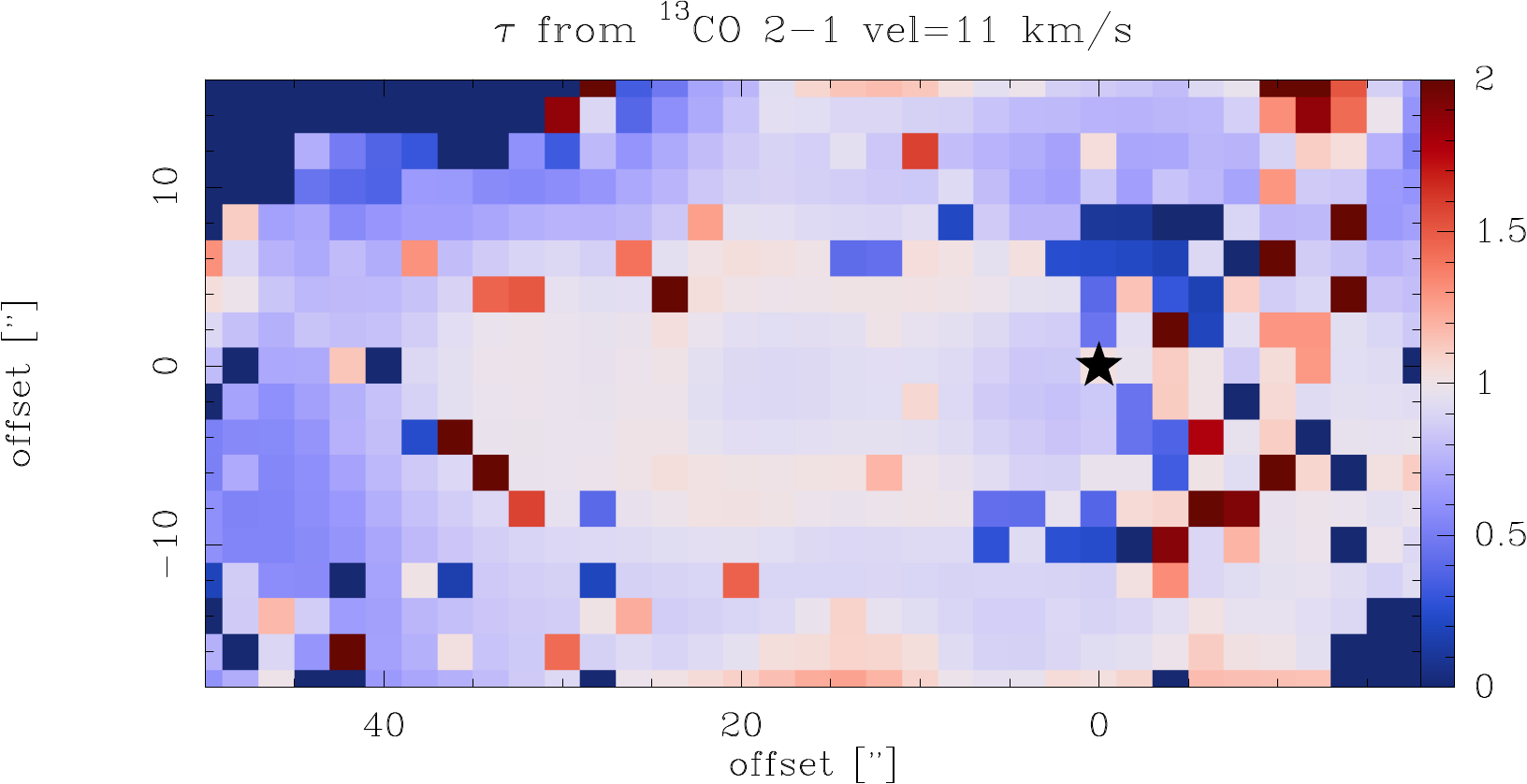} 
\includegraphics [width=8.3cm, angle={0}]{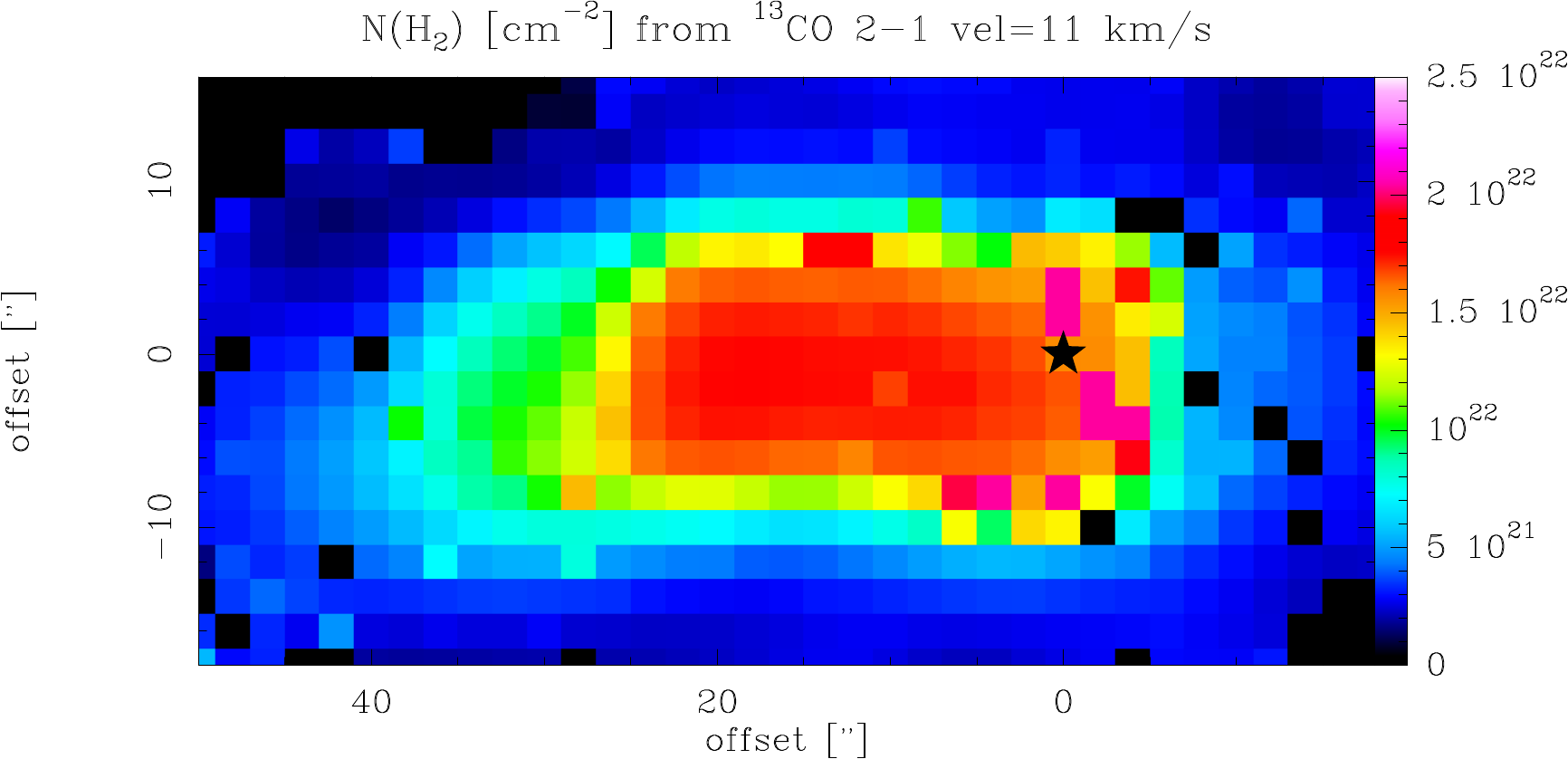} 
\caption{ Maps of physical properties of proplyd \#7. Top: Map of the
  excitation temperature derived from the $^{12}$CO 2$\to$1
  line. Middle: Map of the $^{13}$CO opacity. Bottom: Map of the H$_2$
  column density, calculated from the $^{13}$CO column density. See
  text for all calculations.  }
\label{fig:prop}
\end{center} 
\end{figure}

\subsection{Column density and mass from CO data} \label{appendix-cal-co}
For the calculation of the molecular gas properties such as excitation
temperature, opacity, and column density (see \cite{Mangum2015} and
\citet{Schneider2016b} for details), we assume local thermal
equilibrium (LTE). In this case, the excitation temperature ${\rm
  T}_{ex}(^{12}{\rm CO})$ [K] can be derived from the main beam
brightness temperature ${\rm T}_{mb}(^{12}{\rm CO})$ [K], assuming
this line is optically thick. This is a valid approximation because
the $^{12}$CO/$^{13}$CO 2$\to$1 line ratio in proplyd \#7 is typically
2-10 and thus not reflecting the interstellar $^{12}$C/$^{13}$C
abundance of around 60-70 \citep{Langer1990,Langer1993,Milam2005}. The
excitation temperature then calculates with
\begin{eqnarray} \label{eq:texco} 
{\rm T}_{ex}\left(^{12}{\rm CO}\right) & = & \frac{11.06}{\ln(1+11.06/({\rm T}_{mb}(^{12}{\rm CO})+0.19))}  
.\end{eqnarray}
The $^{13}$CO opacity is estimated using the excitation temperature T$_{ex}$ obtained from $^{12}$CO and the observed $^{13}$CO main beam brightness temperature T$_{mb}$($^{13}$CO) [K] with
\begin{eqnarray} \label{eq:tau13co} 
  \tau_{13}  =  -\ln\left(1-\frac{{\rm T}_{mb}(^{13}{\rm CO})}{(10.58/(e^{10.58/{\rm T}_{ex}}-1)-0.21)}\right). 
\end{eqnarray}
The pixel-by-pixel maps of T$_{ex}$ and $\tau_{13}$ are displayed in
Fig. \ref{fig:prop}. It becomes obvious that the excitation
temperature increases from around 10 K in the outskirts of proplyd \#7
to a maximum value of 35 K in the central regions. Interestingly, the
peak temperatures are not found at the position of the YSO but further
east. The opacity map indicates that the $^{13}$CO line is mostly
optically thin with values below one across proplyd \#7.

The beam averaged total column density $N$ of the optically thin $^{13}$CO molecule can be
determined from the observed line integrated main beam brightness temperature T$_{mb}$($^{13}$CO) with
\begin{eqnarray} \label{eq:N13co} 
N_{\rm ^{13}CO} & = & f(T_{ex}) \int {\rm T}_{mb}(^{13}{\rm CO}) d{\rm v} ,  
\end{eqnarray} 
with   
\begin{eqnarray} \label{eq:ftex} 
f(T_{ex}) & = & \frac{3hZ}{8 \pi^3 \mu_d^2 J_u} 
\frac{e^{E_{up}/kT_{ex}}}{[1-e^{-h\nu/kT_{ex}}] (J(T_{ex}) - J(T_{BG}))}  ,   
\label{ncol}   
\end{eqnarray} 
and 
\begin{eqnarray} \label{eq:J} 
J(T_{ex}) & = & \frac{h\nu}{k(e^{h\nu/kT_{ex}}-1)}  
,\end{eqnarray} 
and $J(T_{BG}) = J(2.7K)$. $h$ and $k$ denote the Planck and the Boltzman constants,
respectively, $E_{up}$ is the energy of the upper level, $\nu$ is the
frequency [GHz], $\mu_d$ is the dipole moment [Debye], $J_u$ is the
upper value of the rotational quantum number and $\int {\rm T}_{mb}(^{13}CO)\,d{\rm v}$ is the velocity integrated line intensity. 

The first two terms of the rotational partition function are given by  
\begin{eqnarray} \label{eq:Zco} 
Z & = & \frac{kT_{ex}}{h B}+1/3  
,\end{eqnarray} 
with the rotational constant B expressed to first order as $\nu$=2BJ$_u$. 

For the determination of the H$_2$ column density $N({\rm H}_2)$, we
use a [$^{12}$C]/[$^{13}$C] abundance of 70 \citep{Langer1990}, and a
[$^{12}$CO]/[H$_2$] abundance of 8.5 10$^{-5}$ \citep{Tielens2010}. We
additionally apply a correction to the hydrogen mass of a factor of
1.36 to account for helium and other heavy elements. The map of the
H$_2$ column density is displayed in Fig. \ref{fig:prop} and indicates
typical values of molecular clouds or clumps in the range of a few
10$^{21}$ cm$^{-2}$ up to 2$\times$10$^{22}$ cm$^{-2}$. These values
correspond well to the {\textit{Herschel}} derived column density map
\citep{Schneider2016a}. \\

\noindent The total mass is then derived from 
\begin{eqnarray} \label{eq:massco} 
{\rm M}(_{tot,{\rm CO}}) = N({\rm H}_2) \, m_{\rm H_2} \, A
,\end{eqnarray} with the mass of a hydrogen molecule $m_{\rm H_2}$ and
the area $A$ of proplyd \#7. We obtain a total mass of 16.3 M$_\odot$
which corresponds well to the one of 19.1 M$_\odot$ obtained from a
high angular resolution (18$''$) map of the Cygnus region presented in
\citet{Schneider2022}. An earlier estimation of 47 M$_\odot$ given in
\citet{Schneider2016a} is based on a lower resolution (36$''$) column
density map which is not corrected for fore- and background
contamination (as was done in \citealt{Schneider2022}).

From the area $A$, we determine an equivalent radius $r$ of the
globule with $r=\sqrt{A/\pi}$ and obtain $r$ = 0.22 pc. An estimate of
the density $n$, using the H$_2$ column density and an extent of
proplyd \#7 considering a spherical geometry with $n = N({\rm H}_2)
/(2\,r)$, yields $n$ = 4.4$\times$10$^3$ cm$^{-3}$.

\subsection{Column density and mass from \CII\ data} \label{appendix-cal-cii}

The \CII\ line profiles do not show a sign of self-absorption (P
Cygnus profiles with a dip at the bulk emission velocity) so that we
assume the most simple scenario of a thermally excited, optically
thin\footnote{We did not detect the $^{13}$\CII\ line that would have
allowed for the $^{12}$\CII\ optical depth to be derived
\citep{Kabanovic2022}.} \CII\ line with atomic hydrogen as the major
collision partner. The \CII\ column density $N_{\rm CII}$ [cm$^{-2}$]
is then calculated \citep{Goldsmith2012} from
\begin{equation}  \label{eq:NII} 
N_{\rm CII}  =  \frac{{\rm I}_{\rm CII}\cdot10^{16}}{3.43} \times \left[ \left(1 + 0.5 \times e^{91.25/{\rm T}_{kin}}\right) \left(1+\frac{2.4 \times 10^{-6}}{C_{ul}}\right) \right]
,\end{equation}
with the line integrated \CII\ emission I$_{\rm CII}$ in [K km s$^{-1}$], the kinetic temperature T$_{kin}$ [K], and the de-excitation rate $C_{ul}$ [s$^{-1}$], which is estimated by
\begin{equation}  \label{eq:Cul} 
C_{\rm ul} =  n \, \times \, R_{ul}    
,\end{equation}
with density $n$ [cm$^{-3}$] and de-excitation rate coefficient $R_{ul}$, derived with  
\begin{equation}  \label{eq:Rul} 
R_{\rm ul} =  7.6 \times 10^{-10} ({\rm T}_{kin}/100)^{0.14}.  
\end{equation}
The atomic mass is then calculated via 
\begin{eqnarray} \label{eq:masscii} 
{\rm M}(_{tot,{\rm H}}) =  N({\rm H}) \, m_{\rm H} \, A  
,\end{eqnarray} 
with the mass $m_{\rm H}$ of a hydrogen atom and assuming an [C]/[H] abundance of 1.6$\times$10$^{-4}$ 
\citep{Sofia2004} for converting $N_{\rm CII}$ column densities into $N_{\rm H}$ 
column densities. We also apply the helium-contribution factor of 1.36.

There are several sources of uncertainty in these calculations. First,
the value of $I_{\rm CII}$ represents only a rough approximation
across the region for which we have \CII\ data (see
Fig.~\ref{spectra}). Second, both the kinetic temperature ($T_{\rm
  kin}$) and the gas density ($n$) must be specified, which presents a
significant challenge.

From the observed maximum main beam brightness temperature of $\sim$9
K (Fig.~\ref{spectra}), we obtain a Rayleigh-Jeans corrected
temperature $T_{RJ}$ with
\begin{equation}  \label{eq:RJ} 
T_{RJ} = \frac{h \nu}{k} \left[ \ln \left(1 + \frac{h \nu}{k T_{mb,C_{II}}}\right) \right] ^{-1} 
\end{equation}
of 38 K. This is the lower limit of the excitation temperature for the
\CII\ line and for thermal excitation T$_{ex} \approx$ T$_{kin}$. We
are approximately in this limit because of the derived densities. The
average density obtained from the CO data analysis is $n = 4.4 \times
10^3$ cm$^{-3}$ while PDR modelling yields local densities of $n \sim
10^5$ cm$^{-3}$ at Position 1 (corresponding to the location of the
YSO) and $n \sim 10^4$ cm$^{-3}$ at Position 2 (in the more quiescent
tail). In all cases, the densities are above the critical density of
3$\times$10$^3$ cm$^{-3}$ for collisions with H atoms.

We calculated the gas mass using various combinations and extreme
values of these parameters. A value of M = 0.7 M$_\odot$ is obtained
for T$_{\rm kin} = 200$ K and $n$ = 10$^4$ cm$^{-3}$, while a mass of
M = 3.9 M$_\odot$ results from T$_{\rm kin} = 38$ K and $n = 4.4
\times 10^3$ cm$^{-3}$. These results suggest that the exact values of
temperature and density have only a moderate influence on the derived
mass. However, we emphasise once again that the \CII\ intensity is
based on a limited number of data points and should be regarded as an
estimate.  Overall, we consider the mass determination to be accurate
within a factor of $\sim$2.

\subsection{Column density and mass from \OI\ data} \label{appendix-cal-oi}

The calculation of the \OI-column density $N_{\rm OI}$ and the
associated mass of the atomic gas is even more challenging to
calculate than for \CII\ because the line often has a high optical
depth
\citep[e.g.][]{Liseau1999,Leurini2015,Schneider2018,Goldsmith2019} and
a more complex line excitation scheme. \OI\ in its ground electronic
state has three fine structure levels that are inverted in energy,
with the $^3P_0$ level having the highest energy and the $^3P_2$ the
lowest. The observable fine structure transitions in the FIR are the
$^3P_0 \to $ $^3P_1$ transition at 145.5 $\mu$m and the $^3P_1 \to$
$^3P_2$ one at 63.2 $\mu$m. We only have data of the 63 $\mu$m line,
which can be excited by collisions with atomic hydrogen at densities
above the critical density of 7.8 10$^5$ cm$^{-3}$
\citep{Goldsmith2019}. The densities in proplyd \#7 are lower (maximum
value is 10$^5$ cm$^{-3}$ from PDR modelling), so that the excitation
will be sub-thermal.

We first assume LTE conditions so that the \OI\ column density is calculated \citep{Leurini2015} with 
\begin{eqnarray} \label{eq:NOI} 
N_{\rm OI} & = & \frac{8 \pi k \nu^2}{h c^3A_{ul}} 
\frac{Z e^{E_{j}/kT_{ex}}}{g_j}
\frac{\tau}{1- e^{-\tau}} 
\int T_{\rm mb,OI} \,d\rm v 
,\end{eqnarray} 
with the Einstein-coefficient $A_{ul}$ = 8.91$\times$10$^{-5}$ s$^{-1}$, $c$ the speed of light, $E_j/k$ =  228 K the upper level energy, $g_j$ = 3, $\tau$ the \OI\ opacity, and $Z$ the partition function that is given by 
\begin{equation}  \label{eq:ZOI} 
Z_{\rm OI} =  \sum_{j=0}^{2} g_j \, e^{-E_j/T_{ex}} = 5 + 3\, e^{-227.7/T_{ex}} + e^{-326.6/T_{ex}}
.\end{equation}
The difficulty now is to to estimate the excitation temperature and the opacity. From the observed main beam brightness temperature of 2 K, we obtain a Rayleigh-Jeans corrected temperature $T_{RJ}$ of 48 K via Eq.~\ref{eq:RJ}, adopted for \OI. 
This is the lower limit for the excitation temperature and in this
case, the line is subthermally excited. Typical excitation
temperatures in bright and dense PDRs are higher \citep{Leurini2015,
  Schneider2018}, of the order of 100 to 200 K, and the \OI\ 63 $\mu$m
line is optically thick and shows self-absorption features
\citep{Boreiko1996, Leurini2015,Schneider2018,
  Goldsmith2021}. However, we do not observe self-reversal in the
\OI\ line profile so we assume that neither the optical depth nor the
excitation temperature are very high in proplyd \#7 (note that the
radiation field is relatively low). Again, we calculate the
\OI\ column density and mass for a representative range of T$_{ex}$
and $\tau$, using a \OI/[H] abundance of 2.5$\times$10$^{-4}$
\citep{Meyer1998} for converting ${\rm N}_{\rm OI}$ column densities
into ${\rm N}_{\rm H}$ column densities. We also apply the
helium-contribution factor of 1.36.

Taking a high T$_{ex}$ of 200 K, we obtain mass values between M =
0.04 M$_\odot$ for $\tau$ = 1 and M = 0.13 M$_\odot$ for $\tau$ = 5,
respectively. The influence of the opacity is thus moderate.  Taking
the lower limit of T$_{ex} = 48$ K and a marginal optically thick
\OI\ line (see below) with $\tau$ = 2, we get a mass of M = 1.8
M$_\odot$.  The \OI\ column density in this case is then 10$^{18}$
cm$^{-2}$ which corresponds well to the results of optically thin or
slightly optically thick, subthermally excited \OI\ emission,
calculated in \citet{Goldsmith2019} using the Molpop–CEP code in a
slab geometry. In their Fig.~2, an excitation temperature of 48 K
corresponds to a density of $\sim$10$^{4.5}$ cm$^{-3}$ and a kinetic
temperature $T_{kin}$ of 100 K.  Figure~4 gives $N_{\rm OI}$ as a
function of optical depth and a value of 1-5 corresponds to a $N_{\rm
  OI}$ $\sim$10$^{18}$ cm$^{-2}$. Figure~8 displays \OI\ line profiles
for $T_{kin}$ = 100 K, $n$=10$^4$ cm$^{-3}$, and a line width of 1.67
km s$^{-1}$ and again, the model maximum antenna temperature of
$\sim$1.7 K for $N_{\rm OI}$ = 10$^{18}$ cm$^{-2}$ corresponds well to
our \OI\ observations for proplyd \#7. Interestingly, the line profile
shows some indications of a flat top, but no significant dip as it is
the case for higher \OI\ column densities. We thus think that our
observed \OI\ line is only marginally optically thick with $\tau$
values of not more than a few.

In addition to the estimates given above, we applied the non-LTE code
RADEX~\citet{vandertak2007} on the two positions where we performed
PDR modelling, namely, position 1 with offset $-4''$, $0''$ and
position 2 with offset $36''$, $0''$. We calculated a parameter grid,
including kinetic temperature T$_{kin}$, H$_2$ density n(H$_2$), and
\OI-column density $N_{\rm OI}$, in a uniform medium. T$_{kin}$ was
varied between 80 and 200 K, n(H$_2$) ranged from $10^3$ to $10^5$
cm$^{-3}$, and $N_{\rm OI}$ from $0.9\times10^{18}$ to
$3\times10^{18}$ cm$^{-2}$. For all three parameters we have
calculated 100 grid points. For T$_{kin}$, we used a linear spacing,
and for n(H$_2$) and $N_{\rm OI}$ logarithmic spacing.

The results are not unique because we cannot unambiguously determine
the densities. However, we still obtain a best-fit solution judging
from the $\chi^2$. For position 1, the excitation temperature is
approximately 49 K, with T$_{kin}$ = 171.7 K, and $n$(H$_2$) =
$8.5\times10^3$~cm$^{-3}$. For position 2, the excitation temperature
is 47 K at T$_{kin}$ = 171.7 K, and $n$(H$_2$) = $6.4\times10^3$
cm$^{-3}$. Both positions yield an optical depth of 1.9 and an
\OI\ column density of $N_{\rm OI}$ = $0.9\times10^{18}$
cm$^{-2}$. All values correspond very well to what we derived from the
LTE approach and by comparing to \citet{Goldsmith2019}.

In summary, the mass estimate has a larger variation depending on
T$_{ex}$, but considering the LTE calculations and the comparison to
\citet{Goldsmith2019}, as well as the RADEX calculations, the most
likely values for the \OI\ column density is $\sim$10$^{18}$ cm$^{-2}$
and the atomic mass is 1.8 M$_\odot$. This value also corresponds well
to the atomic mass of 1-4 M$_\odot$ determined from the \CII\ data.

\end{appendix}
\end{document}